\documentclass[%
reprint, 
superscriptaddress,
 amsmath,amssymb,
 aps, physrev,
]{revtex4-2}

\usepackage[english]{babel}
\usepackage{xcolor}
\usepackage{graphicx}
\usepackage{amsmath}
\usepackage{float}
\usepackage{braket}
\usepackage{algorithm}
\usepackage{algpseudocode}
\usepackage{subfigure}
\usepackage{hyperref}

\begin{document}

\title{Soft information decoding with superconducting qubits}

\preprint{APS/123-QED}

\title{\textbf{Soft information decoding with superconducting qubits} 
}%

\author{Maurice D. Hanisch}
\affiliation{
 IBM Quantum, IBM Research -- Z\"urich, 8803 R\"uschlikon, Switzerland
}%
\affiliation{
 Institute for Theoretical Physics, ETH Z\"urich, 8093 Z\"urich, Switzerland
}%
\author{Bence Het\'enyi}
\affiliation{%
IBM Quantum, IBM Research -- Z\"urich, 8803 R\"uschlikon, Switzerland
}%
\author{James R. Wootton}
\email{james.wootton@unibas.ch}
\affiliation{%
IBM Quantum, IBM Research -- Z\"urich, 8803 R\"uschlikon, Switzerland
}%
\affiliation{%
Center for Quantum Computing and Quantum Coherence (QC2), University of Basel, Petersplatz 1, Basel 4001, Switzerland\\
}%


\date{\today}

\begin{abstract}

Quantum error correction promises a viable path to fault-tolerant computations, enabling exponential error suppression when the device's error rates remain below the protocol's threshold. This threshold, however, strongly depends on the classical method used to decode the syndrome measurements. These classical algorithms traditionally only interpret binary data, ignoring valuable information contained in the complete analog measurement data. In this work, we leverage this richer ``soft information" to decode repetition code experiments implemented on superconducting hardware. We find that ``soft decoding" can raise the threshold by 25\%, yielding up to 30 times lower error rates. Analyzing the trade-off between information volume and decoding performance we show that a single byte of information per measurement suffices to reach optimal decoding. This underscores the effectiveness and practicality of soft decoding on hardware, including in time-sensitive contexts such as real-time decoding.

\end{abstract}

\maketitle

\section{Introduction}

Quantum error correction (QEC) is fundamentally recognized as a central strategy for deploying fault-tolerant computations at scale. It promises exponential error suppression, closing the divide between attainable error rates and those required for practical quantum computing. In recent years, significant progress has been achieved in quantum error correction experiments, demonstrating its emerging capabilities\cite{krinner_realizing_2022, zhao_realization_2022, google_quantum_ai_suppressing_2023, bluvstein_logical_2024, da_silva_demonstration_2024, gupta_encoding_2024, acharya2024quantumerrorcorrectionsurface}.

These advancements are particularly evident in implementing the surface code, a leading candidate for fault-tolerant quantum computation.  Progress has been marked by an upward path from implementing a distance-three ($d=3$) surface code~\cite{krinner_realizing_2022, zhao_realization_2022}, advancing to a distance-five ($d=5$) surface code~\cite{google_quantum_ai_suppressing_2023}, and culminating in a distance-seven ($d=7$) surface code that has also shown evidence for improvements in the fidelities of logical entangling gates with increasing code distance~\cite{bluvstein_logical_2024}. Furthermore, recent experiments claim to have achieved break-even error correction performance, matching or surpassing their physical counterparts. Key milestones include achieving break-even fidelity in the initialization of logical qubits within a color code~\cite{bluvstein_logical_2024}, preparing logical Bell pairs with error rates 800 times lower than their physical level~\cite{da_silva_demonstration_2024}, and successfully distilling magic states with fidelity beyond the break-even point~\cite{gupta_encoding_2024}. Furthermore, in recent experiments, below surface threshold performance was achieved on a distance-seven ($d=7$) code~\cite{acharya2024quantumerrorcorrectionsurface}. These claims highlight quantum error correction's growing effectiveness and practical relevance.

The success of quantum error correction experiments hinges on several critical factors---the choice of quantum code, the error rates of gate operations and mid-circuit measurements, and the performance of the classical decoding algorithm used to process measurement outcomes and determine corrective operations. Usually, outcomes from mid-circuit measurements, inherently analog, are converted into binary data indicating the presence or absence of an error before being analyzed by the decoder. This conversion inevitably results in a loss of information that could degrade decoding accuracy. 

Pattison \textit{et al.} developed a method to circumvent this loss by incorporating the full range of analog data, also known as soft information, into the decoding process~\cite{pattison2021improved} with a weighting algorithm. Their simulations indicated that using soft information in surface code decoding can significantly raise the code's threshold compared to decoders that utilize only binary, or ``hard'', information. This result reveals a substantial improvement in the exponential scaling of error suppression, going beyond just a constant benefit. The findings underscore the importance of utilizing all available information from the experimental setup, thus optimizing quantum error correction strategies.

In addition to the work by Pattison \textit{et al.}, the benefit of employing soft information has been validated through simulations using neural network decoders~\cite{varbanov_neural_2023, bausch_learning_2023}. On the hardware front, soft decoding has been implemented in different settings: for improving the readout fidelity on a Si/SiGe two-qubit system~\cite{xue_repetitive_2020}, and for superconducting qubits using both a distance-three ($d=3$) heavy-hexagon code~\cite{sundaresan_demonstrating_2023} and, recently, a distance-three ($d=3$) two-dimensional repetition code~\cite{ali_reducing_2024}. However, these hardware experiments were limited to at most $d=3$ codes, which restricted their ability to evaluate and compare thresholds between soft and hard decoding---a significant and anticipated aspect of this research.


The experiments of Ref.~\cite{sundaresan_demonstrating_2023} involving $T=10$ rounds of stabilizer measurements managed to post-select for leakage errors---transitions to higher energy states detrimental to quantum error correction~\cite{miao_overcoming_2022}. This post-selection was feasible because the low number of measurement rounds helped restrict the accumulation of leakage, which typically increases with the duration of the experiment. Furthermore, the experiments of Refs.~\cite{ali_reducing_2024} that used Pattison \textit{et al.}'s weighting method did not explicitly treat leakage. Consequently, the issue of leakage and its impact on soft decoding with weighting algorithms, also not addressed in the initial studies by Pattison \textit{et al.} or subsequent simulation efforts, remains a critical area for further investigation.

This work aims to analyze the threshold benefit of soft decoding compared to hard decoding by implementing various distances of the repetition code, ranging from $d=3$ to $d=51$, and analyzing the exponential suppression of logical error rates with code distance for both decoders. Our experiments were conducted on IBM Quantum's 127-qubit device, \texttt{IBM Sherbrooke}. We used this device via the standard IBM Quantum Cloud interface without specific calibrations to improve the accuracy of the specified error correction experiment. Rather than focusing on the lowest logical error rates the hardware is capable of, we sought to compare how incorporating soft information affects the decoding process and improves overall logical error rates. Additionally, given that soft information error correction involves processing and transmitting a greater volume of information, we studied how the performance of soft decoding is influenced when we retain less information from measurements. This analysis addresses the practicality of soft information decoding within a realistic quantum error correction framework such as real-time decoding. Finally, we explore the impact of significantly increasing the number of rounds---up to 100 rounds of stabilizer measurements---to assess how leakage affects the error rates.

\section{Decoding repetition codes with soft information}

\subsection{Repetition codes}

In classical error correction, one method for addressing errors is replicating the data multiple times. This concept extends to quantum error correction, although quantum mechanics requires modifications due to the impossibility of directly cloning quantum states. Our study concerns the distance-$d$ 1D quantum repetition code, where $d$ represents the number of repetitions. A generic quantum state $\ket{\psi} = \alpha \ket{+z} + \beta \ket{-z}$ is encoded in the $Z$-basis as
\begin{align}
    \label{eq:RepCodes_encoding}
    \ket{\psi} \rightarrow \ket{\psi}_L^Z &= \alpha \ket{+z}^{\otimes d} + \beta \ket{-z}^{\otimes d}\\ &= \alpha |00\ldots 0\rangle + \beta |11\ldots 1\rangle,
\end{align}
or in the $X$-basis as 
\begin{align}
    \label{eq:RepCodes_encoding}
    \ket{\psi} \rightarrow \ket{\psi}_L^X &= \alpha \ket{+x}^{\otimes d} + \beta \ket{-x}^{\otimes d}\\ &= \alpha |++\ldots +\rangle + \beta |--\ldots -\rangle.
\end{align}
As can be seen from the encoding, the repetition code only guards against either bit-flip ($X$) or phase-flip ($Z$) errors, but not from both.  Despite not being a ``complete'' quantum code, it is a valuable tool for conceptual demonstrations. They offer a code distance that increases linearly with the number of qubits, making them suitable for high-distance experiments on the intermediate-scale quantum devices that we have today.

We can describe the bit-flip (phase-flip) code by its stabilizers which are $Z$ ($X$) Paulis acting on adjacent qubits
\begin{align}
    S = \langle Z_1Z_2, Z_2Z_3, \ldots, Z_{d-1}Z_d \rangle.
\end{align} 
Consequently, the measurement of stabilizers results in evaluating the parity of neighboring qubits in the $Z$ or $X$ basis. If the parity is odd, an error occured on one of the qubits. 

\begin{figure}
    \centering
    \includegraphics[width=0.9\columnwidth]{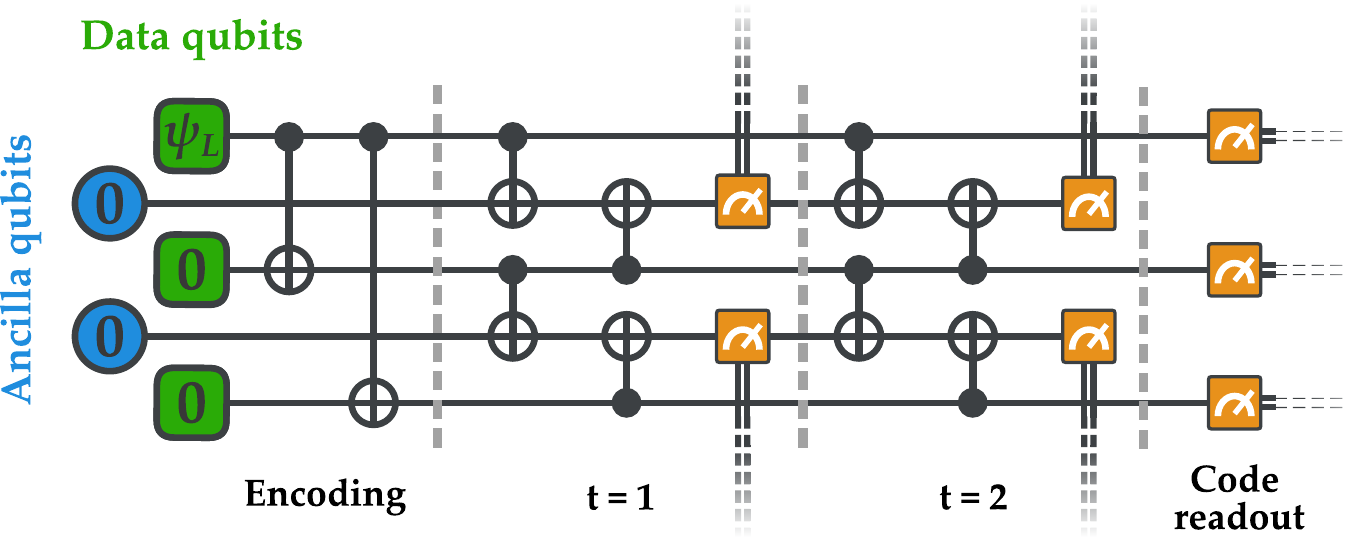}
    \caption{Quantum circuit to implement a repetition code with two rounds of stabilizer measurements on $n=3$ data qubits (orange) and $n=2$ ancilla qubits (blue). The encoding block (E) entangles the logical information in the $X$ or $Z$ basis. The final measurement block (dark green) measures the code qubits in the same basis as the encoding.}
    \label{fig:RepCodes_stabilizer_msmt}
\end{figure}

To implement repetition codes on superconducting hardware, we configure a chain of qubits, utilizing every second qubit as an ancilla to measure the parity between neighboring code qubits. Hence, a repetition code of distance $d$ requires $2d-1$ qubits. We measure the $Z_iZ_{i+1}$ stabilizers by applying a pair of CNOT gates between the data qubits and the ancilla. If an odd number of bit-flips occur on the data qubits, the CNOT gates flip the ancilla. We detect if a flip occurred by measuring the ancilla qubit in the $Z$ basis as shown in Fig.~\ref{fig:RepCodes_stabilizer_msmt}. 

After measuring the stabilizer, one might consider resetting the ancilla to the $|0\rangle$ state to prevent additional error propagation. However, resetting the qubit is a non-trivial task, requiring either waiting for the qubit to decay to the ground state or applying an \textit{active reset} operation. The former would mean that the code qubits in the system must be idle for as long as the ancillas decay, causing them to decay themselves. The latter would require applying an $X$ gate conditioned on the measurement outcome. However, this would require \textit{dynamic operations}, which also cause idling times and specific errors. To circumvent these issues, we do not reset the ancilla qubits.  Moreover, it has been shown that omitting ancilla resets can have benefits in the decoding of memory experiments~\cite{gehér2024resetresetquestion}, further justifying our approach.

Instead of resetting the ancillas physically, we compute the changes in outcomes in the measurement to simulate a scenario as if resets had been performed. Let $\mathbf{m}_t^{\lnot r}$ be the measurement outcomes at time $t$ without resetting the ancilla qubits. We can then write the \textit{corrected} measurement outcomes at time $t$ as
\begin{equation}
    \mathbf{m}_t = \mathbf{m}_t^{\lnot r} \oplus \mathbf{m}_{t-1}^{\lnot r}.
\end{equation}

The quantum circuit for a repetition code for two rounds of stabilizer measurement on $n=3$ data qubits is depicted in Fig.~\ref{fig:RepCodes_stabilizer_msmt}.

\subsection{\label{subsection:decoding}Decoding}

Repetition codes are almost cyclic, enabling their decoding through the minimum weight perfect matching (MWPM) algorithm~\cite{higgott_pymatching_2021, higgott2023sparseblossomcorrectingmillion}. This decoding method is also extensively used for surface codes, widely recognized for their promising potential in fault-tolerant quantum error correction~\cite{krinner_realizing_2022, zhao_realization_2022, google_quantum_ai_suppressing_2023, bluvstein_logical_2024,acharya2024quantumerrorcorrectionsurface}. Therefore, improved decoding performance achieved with repetition codes likely has implications for the broader application with surface codes.

To describe the decoding of our repetition code experiment, we introduce detectors, syndromes, and the decoding graph below. {\it Detectors} measure the parities of measurement results, yielding deterministic values if no errors occur. A {\it syndrome} refers to the detector outcomes, where a 0-entry indicates no error, and a 1-entry indicates the presence of an error. Under circuit-level Pauli noise, the relationship between possible errors and syndromes for a repetition code can be represented as a {\it decoding graph}, where detectors are nodes and potential errors are edges since a single error event is detected by at most two syndromes.

Error chains, i.e., a set of errors with overlapping syndromes, will induce either one syndrome---if the chain terminates at a boundary---or two syndromes. Therefore the decoding problem is equivalent to finding the shortest path on the graph to pair up all the nontrivial syndromes. 
The matching decoder solves this problem on a weighted graph, where the edge weights $w=\log((1-p)/p)$ are calculated using the probability $p$ of the corresponding error event.
The minimum-weight set of edges found by MWPM then corresponds to the most likely set of errors~\cite{higgott_pymatching_2021, higgott2023sparseblossomcorrectingmillion}.

We typically evaluate the merit of a quantum code based on its \textit{threshold value} $p_{th}$, defined as the upper limit of error probability at which the code can still suppress errors. This value originates in the \textit{threshold theorem}~\cite{Nielsen_Chuang_2010}, which posits that assuming a constant probability of a fault $p$ across all potential errors, the logical error rate $\epsilon_L$ decreases exponentially with respect to the \textit{order of fault-tolerance} $f$, which is the maximum number of faults that the code can correct~\cite{ghosh_surface_2012}. For repetition codes with distance $d$, the order of fault-tolerance is $f = \lfloor d/2\rfloor$ as the code can detect up to $d-1$ errors but only unambiguously correct up to $\lfloor d/2\rfloor$ errors. Consequently, the logical error rate $\epsilon_L$ exponentially decreases with the distance of the code as
\begin{equation}
    \label{eq:scaling_e_L_f_p+1}
    \epsilon_L \propto \left(\frac{p}{p_{\text{th}}}\right)^{f+1}\left. \vphantom{\frac{p}{p_{\text{th}}}}\right|_{f = \lfloor d/2\rfloor}.
\end{equation}

The code in question does not solely determine the threshold value, as it is also significantly influenced by the decoding method~\cite{google_quantum_ai_exponential_2021}. Typically, researchers simulate the performance of a code and its decoder across varying error rates to estimate the threshold. Moreover, the error parameters in the circuit-level error model do not need to be equal, requiring a threshold parameter for each error mechanism~\cite{hetenyi2024tailoring}. Determining these thresholds in experiments would require precise manipulation of the noise model and, hence, gates with accurately controllable error rates---a daunting task. 

Instead, we employ the \textit{error suppression factor} $\Lambda$ as a practical metric for comparing the performance of various codes and decoders~\cite{kelly_state_2015}. We define $\Lambda$ through the scaling of the logical error rate $\epsilon_L$ as
\begin{equation}
    \label{eq:logical_error_rate_lambda_factor}
    \epsilon_L \propto \left(\frac{1}{\Lambda}\right)^{\lfloor \frac{d}{2}\rfloor+1}.
\end{equation}
Consequently, $\Lambda$ effectively measures \textit{how far below the threshold the error rate of a system is}, without needing to determine the threshold itself. 

Comparing Eq.~\eqref{eq:logical_error_rate_lambda_factor} with Eq.~\eqref{eq:scaling_e_L_f_p+1}, we see that determining the $\Lambda$-factor for different decoders allows us to effectively compare the threshold as the ratio of the $\Lambda$-factors corresponds to the ratio of the thresholds
\begin{equation}
    \label{eq:threshold_ratio_lambda_factor}
    \frac{p_{th,1}}{p_{th,2}} = \left(\frac{\Lambda_1}{\Lambda_2}\right).
\end{equation}

When decoding multiple rounds of syndrome measurements, we obtain the \textit{total logical error rate} $P_L$ rather than the previously discussed logical error rate \textit{per round} $\epsilon_L$. Assuming each round incurs a logical error with probability $\epsilon_L$, the total logical error rate $P_L$ after $T$ rounds is given by~\cite{google_quantum_ai_exponential_2021}:
\begin{equation}
    P_L = \frac{1}{2}\left(1 - (1-2\epsilon_L)^T\right).
\end{equation}
Using $(1-2\epsilon_L)^T \approx 1-2T\epsilon_L$ for $\epsilon_L \ll 1$, we can approximate the total logical error rate as
\begin{equation}
    P_L \approx T\epsilon_L.
\end{equation}

Consequently, to compute the $\Lambda$-factor when decoding multiple rounds, we must first derive the logical error per round $\epsilon_L$ from the total logical error rate $P_L$ and the number of rounds $T$. We then use Eq.~\eqref{eq:logical_error_rate_lambda_factor} to determine the $\Lambda$-factor. Utilizing this methodology, we will calculate the lambda factors to compare the performance of various decoders in the subsequent sections.

\subsection{Soft information} 

On most quantum devices, the measurement process culminates in a final continuous signal. We then discretize this analog signal into a binary value indicating the outcome of the measurement. This discrimination process leads to \textit{loss of information} and \textit{assignment errors}. To describe these errors, we discuss readout on superconducting hardware, which is the focus of our experiments. However, the principles are broadly applicable across various quantum computing platforms, as discussed later.

\begin{figure}
    \centering
    \includegraphics[width=0.9\columnwidth]{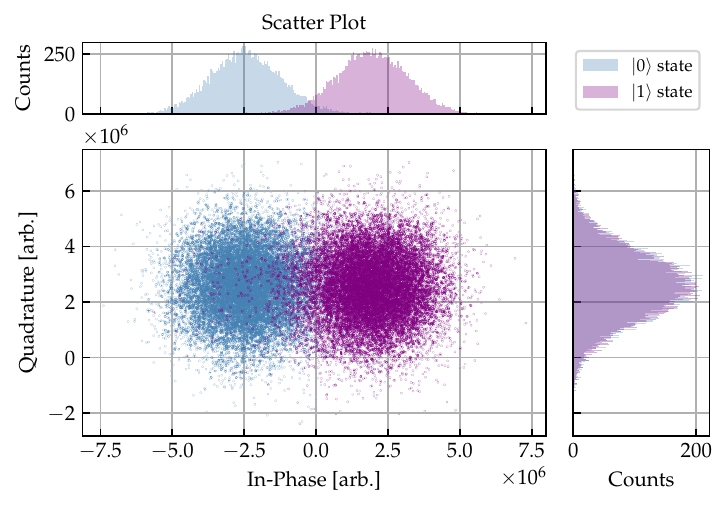}
    \caption{IQ data for qubit 72 of the \texttt{IBM Sherbrooke} device prepared in the $\ket{0}$ (blue) and $\ket{1}$ (violet) state. The data shows two distinct Gaussian distributions corresponding to the qubit's computational states.}
\label{fig:iq_data_example}
\end{figure}

In superconducting hardware, we perform qubit readouts by coupling a resonator to the qubit. The qubit's state then induces a \textit{dispersive shift} in the resonator's frequency. To detect this shift, we apply a microwave tone to the resonator and measure the reflected signal, which has an altered amplitude and phase according to the qubit's state. We integrate this reflected signal to extract two key metrics: the \textit{in-phase} ($I$) and \textit{quadrature} ($Q$) components and plot them on the \textit{IQ plane}~\cite{bronn_fast_2017}. This plane has $I$ and $Q$ on its axes and visually represents the qubit's state through the distribution of $I$ and $Q$ points. Statistically, the $I$ and $Q$ points cluster into two distinct Gaussian distributions on the IQ plane, each representing one of the qubit's possible states. Fig.~\ref{fig:iq_data_example} shows statistical IQ data obtained from a superconducting qubit. To ascertain the qubit's state from the reflected signal, we assess the likelihood of a data point belonging to each Gaussian and convert it into a binary outcome. The dispersive shift determines the separation between the two Gaussians, whereas the width of the Gaussian curves is dependent, among other things, on the duration of the integration window. Despite efforts to separate these distributions, a non-zero overlap exists, creating a situation where an IQ data point might fall into an area more likely associated with the incorrect state. Such occurrences lead to a \textit{misassignment error}, commonly called a \textit{soft flip}.

We will use the following model to describe these misassignments in a device-independent manner. We assume that during the measurement the qubit state $\ket{\psi}$ gets projected onto a \textit{true state} $\ket{\bar{z}} \in \{\ket 0,\ket 1\}$ with probability $\mathbb{P}[\bar{z}] = |\braket{\bar{z}|\psi}|^2$. We then observe a \textit{soft outcome} $\mu$ according to the probability distribution $f^{\bar{z}}(\mu) = \mathbb{P}[\mu \mid \bar{z}]$, where $\mathbb{P}[\mu \mid \bar{z}]$ is the conditional probability of measuring $\mu$ provided that the state was projected onto $\bar{z}$. To classify the qubit state and obtain the \textit{estimated state} $\hat{z}$, we compare the probabilities of observing the soft outcome for each state and use a \textit{maximum likelihood assignment}:
\begin{equation}
    \label{eq:maximum_likelihood_assignment}
    \hat{z}= \begin{cases}0 & \text { if } f^{(0)}(\mu) \geq f^{(1)}(\mu) \\ 1 & \text { otherwise }\end{cases}.
\end{equation}

We summarize the estimated stabilizer outcomes at round $t$ in an \textit{outcome vector} $\mathbf{\hat{z}}_t$ which either corresponds to $\mathbf{m}_t$ or $\mathbf{m}_t^{\lnot r}$, depending on whether we reset the ancilla qubits or not.

With this model, a \textit{soft flip} occurs if the estimated state is not equal to the true state, $\hat{z}\neq\bar{z}$, meaning that we wrongfully estimated the state of the qubit from the soft outcome $\mu$. On the other hand, we call a \textit{hard flip} an error that changes the true state $\bar{z}$ of the qubit before or during the measurement.

The probability of a \textit{soft flip} given an observed soft outcome $\mu$ can be written as 
\begin{equation}
    \mathbb{P}[\text{soft} \mid \mu] = \frac{\mathbb{P}[\text{soft}, \mu]}{\mathbb{P}[\mu]}, 
\end{equation}
where
\begin{equation}
    \label{eq:soft_and_mu_probability}
    \mathbb{P}[\text{soft}, \mu] = \begin{cases} {f^{(1)}(\mu)}\cdot\mathbb{P}[\bar{z}=1] & \text { if } f^{(0)}(\mu) \geq f^{(1)}(\mu) \\ {f^{(0)}(\mu)\cdot\mathbb{P}[\bar{z}=0]} & \text { otherwise,}  \end{cases}
\end{equation}
and 
\begin{equation}
    \mathbb{P}[\mu] = f^{(0)}(\mu)\cdot\mathbb{P}[\bar{z}=0] + f^{(1)}(\mu)\cdot\mathbb{P}[\bar{z}=1].
\end{equation}

\begin{figure}
    \centering
    \includegraphics[width=0.95\columnwidth]{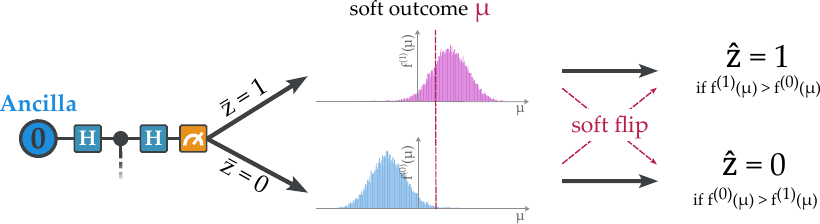}
    \caption{Illustration of the measurement process model. The qubit state gets projected onto the eigenspace of the true state $\bar{z}$, leading to a soft outcome $\mu$ according to the probability distribution $f^{\bar{z}}(\mu)$. We then obtain the estimated state $\hat{z}$ through a maximum likelihood assignment.}
    \label{fig:measurement_model}
\end{figure}

Consequently, if we observe a soft outcome $\mu$ and we estimate the state to be $\hat{z}$, the probability of a soft flip is
\begin{equation}
    \label{eq:soft_flip_probability_priors}
    \mathbb{P}[\text{soft} \mid \mu] =  \left[1+\frac{\mathbb{P}[\bar{z}=\hat{z}]}{\mathbb{P}[\bar{z}=\hat{z}\oplus 1]}\frac{f^{(\hat z)}(\mu)}{f^{(\hat z\oplus1)}(\mu)}\right]^{-1},
\end{equation}
where $\oplus$ denotes the addition modulo-2. This model allows us to describe the soft flips device independently, only requiring the distributions $f^{(0)}(\mu)$ and $f^{(1)}(\mu)$ to be known. An illustration of the measurement process model is shown in Fig.~\ref{fig:measurement_model}.

The read-out is typically calibrated by preparing a state $\ket{0}\text{ or }\ket{1}$ and minimizing the probability to measure a $\ket{1} \text{ or } \ket{0}$ respectively. This doesn't distinguish between hard and soft flips. However, it is possible to calibrate by minimizing a figure of merit that captures the ambiguity in the distributions of the analog signal \cite{D'Anjou_gen_chernov}. 

In the subsequent discussions, we will assume that the priors are equal, $\mathbb{P}[\bar{z}=0] = \mathbb{P}[\bar{z}=1] = 0.5$, hence leading to 
\begin{equation}
    \label{eq:soft_flip_probability}
    \mathbb{P}[\text{soft} \mid \mu] =  \left[1+\frac{f^{(\hat z)}(\mu)}{f^{(\hat z\oplus1)}(\mu)}\right]^{-1}.
\end{equation}
This assumption generally holds for stabilizer measurements after sufficient rounds, as the ancilla qubits will likely be randomly in the $|0\rangle$ or $|1\rangle$ state. However, for the initial rounds and particularly for the final logical code readout, this assumption may not hold. Therefore, incorporating additional information derived from Pauli tracing~\cite{gidney_stim_2021} could improve the model's accuracy and, consequently, the efficacy of the methods discussed in this chapter.

\subsection{Decoding graph with soft flips}

\begin{figure}
    \centering
    \includegraphics[width=0.9\columnwidth]{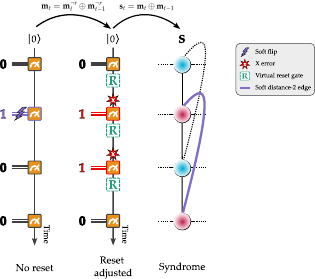}
    \caption{Timeline of measurement outcomes without resets, the inferred corrected outcomes, and its corresponding single stabilizer column of the decoding graph $G_T$. A soft flip occurs at time t=2, and the qubit state is wrongfully read out as $\ket{1}$. When calculating the reset adjusted measurement outcomes, the soft flip is equivalent to two consecutive hard flips. A hard flip can be seen as a bit-flip error on the ancilla before measurement. Due to this overcorrection, the column contains a pair of flipped syndromes separated by two time steps.}
    \label{fig:soft_timeline}
\end{figure}

As discussed, hard flips alter the ancilla qubit's state, leading to a faulty measurement outcome. In contrast, soft flips only change the measurement outcome; therefore, the outcomes disagree with the state of the ancilla. This difference is crucial when we do not reset the ancilla qubits. When correcting the measurement outcomes to synthesize ``reset outcomes'' $\mathbf{m}_t = \mathbf{m}_t^{\lnot r} \oplus \mathbf{m}_{t-1}^{\lnot r}$, we assume the ancilla qubits are in the state we measured them in. This assumption is valid for hard flips but not soft flips. Consequently, if an outcome gets altered due to a misassignment, we will ``overcorrect,'' leading to a pair of syndromes separated by \textit{two} time steps, as illustrated in Fig.~\ref{fig:soft_timeline}.

Consequently, a \textit{single error channel}---a soft flip---can connect two syndrome nodes separated by two time steps. We represent this error channel in the decoding graph by adding a \textit{distance-two edge}. We will call this distance-two edge the \textit{soft edge}. Moreover, as a soft flip will only flip pairs of syndromes, we can still decode by matching. It is important to note that this \textit{weight-two-time error} has implications on the approximation $T\gg d$ to an optimal \textit{infinite memory}. Nonetheless, we see negligible effects on the decoding performance.

These weight-two-time errors do not occur for all rounds; a discrepancy arises when considering the last round of stabilizers. In a memory experiment, we artificially compute the final round of stabilizer outcomes from the code measurements. This last round is, therefore, insensitive to soft-flips on the ancillas. I.e., a soft flip in the previous round of stabilizer measurements would only cause one faulty measurement outcome, not two consecutive ones. Hence, it produces a pair of flipped syndromes separated by a single time step. A hard flip has the same effect. Consequently, soft and hard flips are two independent, indistinguishable error channels for the last round. Their probabilities combine to give the total probability of a measurement error in the last round as
\begin{equation}
    \label{eq:soft_and_hard_probability_last_round}
    \mathbb{P}_E[\mathtt{E_m}] = p_h(1-p_s) + (1-p_h)p_s,
\end{equation}
with $p_h = \mathbb{P}_E[\mathtt{E_h}]$ and $p_s = \mathbb{P}_E[\mathtt{E_s}]$ the probabilities of a hard and soft flip, respectively. 

On the other hand, when we \textit{do reset} the ancilla qubits, these weight-two-time errors should not appear in theory. However, an active reset with an $X$ gate conditioned on the stabilizer measurement outcome will be faulty when a soft flip occurs. This faulty reset would then have the same effect as a faulty artificial correction $\mathbf{m}_t = \mathbf{m}_t^{\lnot r} \oplus \mathbf{m}_{t-1}^{\lnot r}$ from the no-reset case. One could circumvent this issue using a \textit{second independent measurement} for the active resets.

\begin{figure}
    \centering
    \includegraphics[width=0.5\columnwidth]{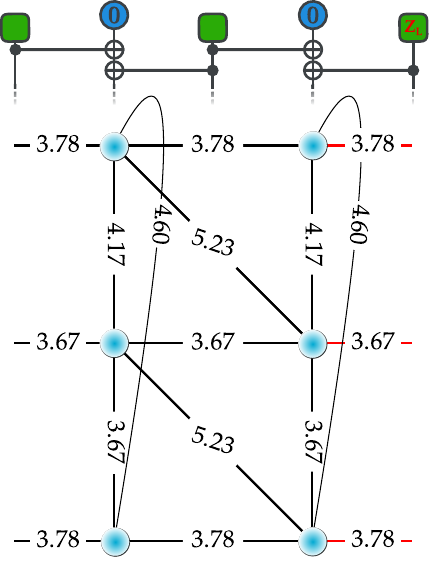}
    \caption{Decoding graph for a repetition code of distance $d=3$ with $T=2$ stabilizer round, including the final code qubit readout. The graph represents the syndrome nodes (blue circles) connected by code qubit and measurement error edges. The weights of the edges are typical for the \texttt{IBM Sherbrooke} noise model. The red edges represent the qubit that carries the logical information.}
    \label{fig:decoding_graph}
\end{figure}

The repetition code is a one-dimensional code with an interleaved arrangement of qubits and stabilizers along a line. This layout forms a two-dimensional decoding graph, where a line of syndrome nodes connected by code qubit edges extends in the time direction. The hard flip channel connects consecutive syndrome nodes, while the soft flip channel creates a distance-two time edge. Data qubit error channels connect neighboring nodes. An additional error channel links the diagonal syndromes; this error is caused by an error occurring between CNOTs during the stabilizer measurement. This channel's orientation depends on the error propagation direction, determined by the CNOT gate ordering. An example of the decoding graph for a repetition code with $d=3$ is shown in Fig.~\ref{fig:decoding_graph}.

\subsection{Dynamic weighting using soft information}

In this section, we will explore how we can improve decoding by using information contained in the analog nature of the signal. 

A simple approach to improve decoding with soft information consists of first using calibration data to estimate the distributions $f^{(0)}(\mu)$ and $f^{(1)}(\mu)$. We then utilize these distributions to estimate the mean probability of a soft flip according to Eq.~\eqref{eq:soft_and_mu_probability} as
\begin{align}
    \mathbb{P}[\text{soft}] &= \int \mathbb{P}[\text{soft}, \mu] \, \mathrm{d}\mu \\
    &= \sum_{z=0}^1\int_{f^{(z)}(\mu) < f^{(z \oplus 1)}(\mu)} f^{(z)}(\mu) \, \mathrm{d}\mu.
\end{align}
We can then augment the decoding graph with soft edges weighted by $p=\mathbb{P}[\text{soft}]$. This method enables a better-fitted noise model, enhancing the decoding performance.

However, we can approach utilizing soft information from an alternative angle and \textit{dynamically adjust the weights} in the decoding graph. Instead of utilizing the estimated states of the stabilizer measurements $\{\mathbf{\hat{z}_t}\}_{t=1, \ldots, T}$ as our input for the decoder, we retrieve the soft outcomes $\{\boldsymbol{\mu}_t\}_{t=1, \ldots, T}$ before classification. Employing these soft outcomes alongside estimations of the probability distributions $f^{(0)}(\mu)$ and $f^{(1)}(\mu)$, we compute the probability of a soft flip $\mathbb{P}[\text{soft} \mid \mu]$ for each measurement outcome $\mu$, as per Eq.~\eqref{eq:soft_flip_probability}. This calculation enables us to dynamically adjust the weights in the decoding graph based on the measurement outcome's ambiguity, thus refining our noise model and decoding graph on a \textit{per-shot basis}. The full procedure is outlined in Algorithm \ref{algorithm:soft_decoding}.

Using the soft outcomes directly is unique as it is the only additional information we can extract immediately from the stabilizer rounds. This extra information can remedy the loss of information from classification. Moreover, this approach is also largely independent of the noise model, establishing it as a general approach for enhancing the decoding of graph-like structures with soft information.

\subsection{\label{subsection:leakage}Soft decoding in the presence of leakage}

Transmon qubits, featured in IBM's devices, are a leading platform for quantum computation. However, transmons have a finite anharmonicity, allowing transitions to higher energy states. Such transitions can excite a qubit, resulting in \textit{leakage errors}. These excitations often occur when implementing two-qubit gates~\cite{motzoi_simple_2009, chen_measuring_2016}, as well as measurements~\cite{barends_superconducting_2014, yan_tunable_2018, rol_fast_2019, negirneac_high-fidelity_2021}.

Leakage poses a significant challenge in quantum error correction. Once a qubit leaks, it does not operate as intended, leading to persistent faults over multiple stabilizer measurements. Quantum error correction typically assumes that errors are uncorrelated; however, leakage undermines this assumption, posing a significant challenge for the experimental implementation of quantum error correction protocols. Miao \textit{et al.} demonstrated the detrimental effect of leakage on decoding performance in superconducting qubits~\cite{miao_overcoming_2022}, showing a much higher sensitivity of code performance to leakage than Pauli errors. Past experiments involving multiple rounds of quantum error correction have focused on identifying and post-selecting leaked qubits, as seen in the studies conducted by 
\cite{google_quantum_ai_exponential_2021, krinner_realizing_2022}.

Soft information decoding experiments have also selectively removed samples with detected leaked qubits~\cite{ali_reducing_2024, sundaresan_demonstrating_2023}. Notably, these studies involved no more than $T=16$ rounds of measurements. In contrast, our experiments extend to $T=100$ rounds, significantly increasing the likelihood of leakage and thus reducing the yield to nearly zero if post-selection were applied.

Several strategies have been developed to mitigate leakage in quantum systems, such as implementing leakage-reducing reset protocols~\cite{google_quantum_ai_exponential_2021,mcewen_removing_2021}, applying optimal control techniques to gate operations~\cite{werninghaus_leakage_2021}, and integrating leakage reduction units directly into the circuit design~\cite{suchara_leakage_2014, fowler_coping_2013}. In line with our decision to use IBM Quantum’s deployed device \texttt{IBM Sherbrooke}, which is not optimized for quantum error correction, we explored the extreme case where no leakage mitigation is applied at the hardware level. Instead of focusing on implementing sophisticated leakage-reduction protocols, we investigated how effectively leakage can be managed solely through classical decoding techniques. As Fowler \textit{et al.}~\cite{fowler_coping_2013} showed, a threshold still exists under leakage, although significantly lower, making this a compelling test of soft decoding's robustness under realistic, unmitigated conditions.

Leakage manifests as a separate Gaussian distribution in the IQ plane, typically positioned between the Gaussians representing the computational states \cite{sundaresan_demonstrating_2023}. This separation allows us to detect leakage using the IQ data already available, enabling the classification of leaked qubits based on their IQ position.

We discovered that leakage can be addressed within the same framework used for handling ambiguities, namely through reweighting the decoding graph. If an ancilla qubit leaks, its soft outcome $\mu$ becomes uninformative about the stabilizer's eigenvalue it is associated with. Hence, we declare the measurement outcome maximally ambiguous and set the corresponding distance-two time edge to zero. This allows the decoder to effectively disregard the measurement outcome by including a zero-weight edge in the matching.

To maintain compatibility with the soft decoding framework, we use the $\ket{0}$-$\ket{1}$ discriminator to assign a binary value to the stabilizer outcome. However, since the leakage Gaussian typically lies between the Gaussians for the computational states, the attributed value will be randomly assigned as 0 or 1. Consequently, the outcome $\hat{z}$ is independent of the ancilla qubit's actual state, meaning leakage does not act as a standard hard flip error but rather a bare misassignment. Consequently, we do not set the distance-one time edge weight to zero when a leaked IQ point is measured.

\begin{figure}[!ht]
    \centering
    \includegraphics[width=0.7\columnwidth]{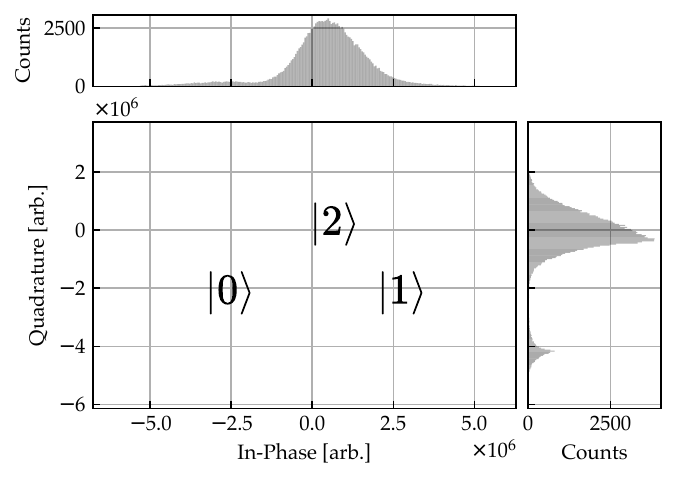}
    \caption{Vizualization of outliers in the IQ plane according to our filtering method for qubit 3 of \texttt{IBM Sherbrooke}. Red points indicate outliers, which are classified as leaked states. They are determined based on their sampling probabilities from the distribution $f^{(z)}(\mu)$, with points falling below the 1\% threshold considered as potential leaked states.}
    \label{fig:outliers_iq}
\end{figure}

To determine whether an IQ point is leaked, we repurposed calibration data not originally intended for leakage detection. Retrieving the calibration data involves a simple circuit with just state preparation and subsequent readout. Ideally, this circuit should not exhibit leakage, although this assumption is only approximate for our experiments. For each IQ point $\mu$, we assess it against the calibration data to determine whether it qualifies as an \textit{outlier}. We define an outlier as a point with a sampling probability from $f^{(z)}(\mu)$ below a specified threshold for both states $z=0,1$. We have empirically set this threshold at 1\% to optimize detection efficacy. We consider such points as leaked states and assign them a soft flip probability $p_s$ of 0.5. We display the effects of this method on our data in Fig.~\ref{fig:outliers_iq}.

This method encounters classification errors in identifying leakage for two primary reasons. 

First, it is challenging to distinguish genuine outliers of computational states from those of leaked states. The leaked state distribution often lies close to the computational state distributions, making it necessary to use a conservative threshold to ensure leaked states are correctly identified. This approach, however, increases the likelihood of misassigning legitimate $\ket{0}/\ket{1}$ outliers as leaked state. For instance, if an IQ point from a ``good'' computational state lies a few standard deviations away from the mean (e.g. $I \leq 5\times10^6$), it might be erroneously classified as an outlier and, thus, as leakage.

Second, as described in~\cite{sundaresan_demonstrating_2023}, the Gaussian distribution of the leaked state has a variance similar to the computational states, ideally calling for a linear decision boundary between leaked and computational states. Our approach, based on outlier detection, results in a circular decision boundary, which tends to underestimate the number of leaked states, as shown in Fig.~\ref{fig:outliers_iq}. Addressing this limitation would ideally involve including a distinct preparation of leaked states in the calibration data and performing a three-state discrimination process, as described in~\cite{sundaresan_demonstrating_2023}.

We attempt to balance these errors by setting the threshold at 1\%.

Despite these imperfections, it still significantly enhances decoding performance, as we will demonstrate in subsequent sections. However, our current approach to handling leakage lacks a robust theoretical basis, and the information indicating a state leaked might be more effectively utilized in a specialized decoder than by simply assigning a soft flip probability of 0.5.

\section{Results} 

\subsection{\label{subsection:T50_simulations}Simulations}

To explore the advantages of soft decoding repetition codes on superconducting hardware, we initially performed simulations that mimic the behavior of realistic quantum hardware to assess the effectiveness of soft decoding. After these preliminary simulations, we conducted experiments on IBM Quantum's deployed device \texttt{IBM Sherbrooke}, a 127-qubit Eagle device not specifically tuned for quantum error correction tasks like repetition codes. Rather than mitigate leakage and other measurement errors through fine-tuning of the measurement process, as would be done in a full demonstration of an error correcting code, we instead simply use the measurements as provided ``out-of-the-box" through Qiskit~\cite{javadiabhari2024quantumcomputingqiskit}. This allows us to investigate how much the leakage can be mitigated purely through the decoding. Using this general-purpose approach, we did not aim to maximize QEC performance but to evaluate how soft decoding behaves under practical conditions comparing the outcomes with those of the simulations.

We investigate the logical error rate over $T=50$ rounds of stabilizer measurements, simulating a repetition code of distance $d=51$. This is the maximal feasible odd distance that we found while ensuring the used line of qubits did not include any CNOT connection with fidelity $<90\%$ and ancilla with readout fidelity $<80\%$. Additionally, we derive results for all odd distances ranging from $d=3$ to $d=51$, employing the subsampling method utilized in our hardware experiments, as described in Appendix \ref{appendix:experimental_setup}. This approach allows us to closely mimic the hardware conditions and evaluate the performance of soft decoding across various distances. We compare two decoding strategies: \textit{Dynamic Soft Decoding} and \textit{Calibrated Hard Decoding}, where the latter estimates each error probability from calibration data, as described in Appendix \ref{appendix:noise_model}, and weights the decoding graph once. 

While this setup with $T=50$ does not approach the condition of $T \gg d$, it aligns with the conventional experimental practice where $d$ is set equal to $T$, ensuring that the ``time distance'' equals the code distance. Ref.~\cite{google_quantum_ai_exponential_2021} showed that time boundary effects in repetition codes do not affect the logical error rate per round for $T \geq 10$, supporting our decision to set $T=50$.

Furthermore, because we modeled symmetric decoherence channels, the noise model is independent of the prepared logical state. Consequently, we only simulate one of the logical states for each basis.

\begin{figure}
    \centering
    \subfigure[Prepared logical state $\ket{\psi}_L = \ket{+z}_L$.]{
        \includegraphics[width=0.46\columnwidth]{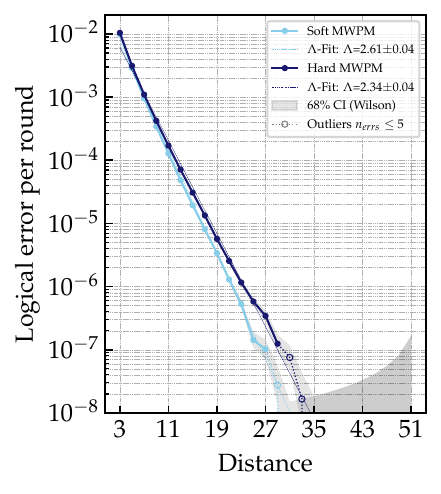}
        \label{fig:SIM_Z_1_eL}
    }
    \hfill 
    \subfigure[Prepared logical state $\ket{\psi}_L = \ket{+x}_L$.]{
        \includegraphics[width=0.46\columnwidth]{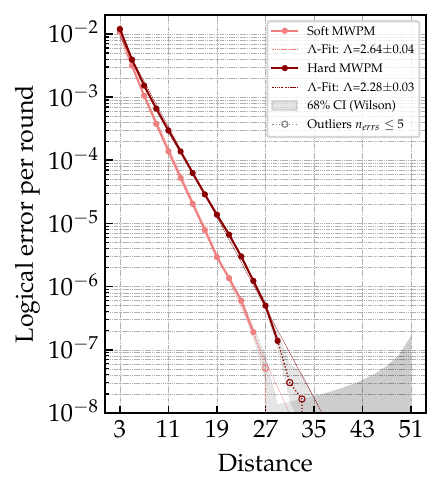}
        \label{fig:SIM_X_1_eL}
    }

    \caption{Simulation results of the \texttt{IBM Sherbrooke} device, depicting the logical errors per round as a function of code distance for the soft and hard MPWM decoders for $T=50$ rounds of stabilizer measurements. The Lambda factors $\Lambda$ are fitted from the curves' slopes in the logarithmic scale.} 
    \label{fig:SIM_combined_eL_1}
\end{figure}

Simulation results for soft decoding of repetition codes are illustrated in Fig.~\ref{fig:SIM_combined_eL_1} for the $Z$ and $X$ bases. These plots show the \textit{logical error per round} $\epsilon_L$ and fit the Lambda factor $\Lambda$ from the curve's slope in logarithmic scale as a function of varying code distances from $d=3$ to $d=51$ for $T=50$ rounds of stabilizer measurements. They compare the soft and hard decoders and include the 68\% confidence intervals calculated using the Wilson score interval~\cite{wilson_probable_1927}. The confidence intervals reflect the varying number of shots for different distances, with higher distances having fewer shots due to subsampling, as indicated by the intervals' width. Points with fewer than \textit{five} logical error events appear as dashed lines, suggesting insufficient statistical data to represent the logical error rate.

Importantly, we lack data points for distances $d \geq 35$, as the error rates drop below $10^{-7}$, necessitating more than tens of millions of shots to record a single logical error event. However, processing even one million shots poses a significant computational challenge.

Both error rates demonstrate a linear decay with increasing code distance on a logarithmic scale, indicating exponential suppression of errors, as detailed in Section \ref{subsection:decoding}. The curve for the soft decoder consistently lies below that of the hard decoder across all distances, demonstrating the higher performance of the soft decoder. Notably, the error rates for the maximal distances resolved for the $Z$ and $X$ bases are \textit{three} and \textit{fifteen times lower}, respectively.  

Fig.~\ref{fig:SIM_combined_eL_1} demonstrates that the soft decoder's curve has a steeper slope than that of the hard decoder, indicating a higher threshold and an exponentially superior performance for the soft decoder. As detailed in Section \ref{subsection:decoding}, the Lambda factor quantifies how many times lower the physical error rates are compared to the decoder's threshold. Consequently, the ratio of the Lambda factors between the soft and hard decoders directly corresponds to the ratio of their thresholds. By comparing the two $\Lambda$-factors, we determine that the threshold of the soft decoder surpasses that of the hard decoder by $11.5\%$ for the $Z$ basis and $15.7\%$ for the $X$ basis, respectively. These improvements exceed the bounds of statistical noise, confirming the significant exponential enhancement provided by soft decoding for repetition codes.

\subsection{\label{subsection:Results_50_rounds}Decoding data from quantum hardware}

While the simulations offer valuable insights, they do not fully capture complexities such as correlated errors, including qubit crosstalk and leakage, which can affect hardware performance. Although recent research has proposed more accurate metrics for closer-to-hardware simulations using calibrated device error rates~\cite{hesner2024usingdetectorlikelihoodbenchmarking}, we did not pursue this approach in our current study. Therefore, to better understand the practical implications of soft decoding, we must proceed with experimental implementation on the actual hardware.

\begin{figure}
    \centering
    \subfigure[Prepared logical state $\ket{\psi}_L = \ket{+z}_L$.]{
        \includegraphics[width=0.46\columnwidth]{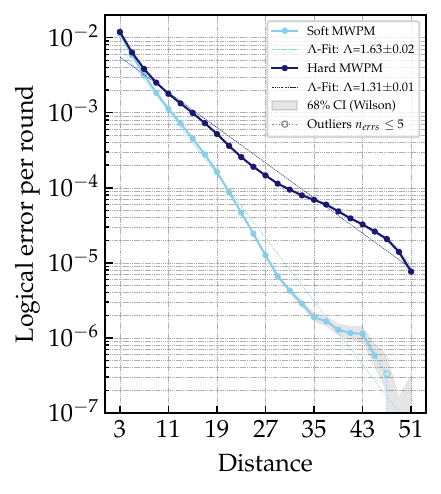}
        \label{fig:RepCodes_soft_Z0_lambda}
    }
    \hfill 
    \subfigure[Prepared logical state $\ket{\psi}_L = \ket{-z}_L$.]{
        \includegraphics[width=0.46\columnwidth]{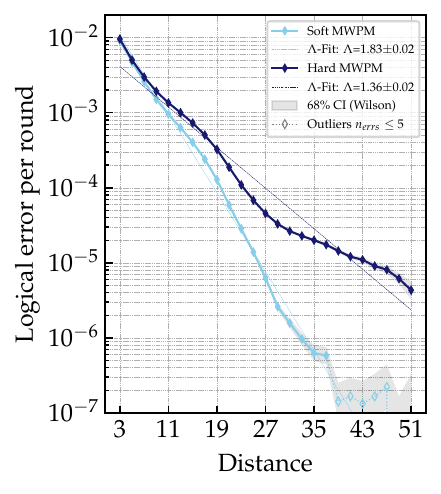}
        \label{fig:RepCodes_soft_Z1_lambda}
    }\\ 

    \subfigure[Prepared logical state $\ket{\psi}_L = \ket{+x}_L$.]{
        \includegraphics[width=0.46\columnwidth]{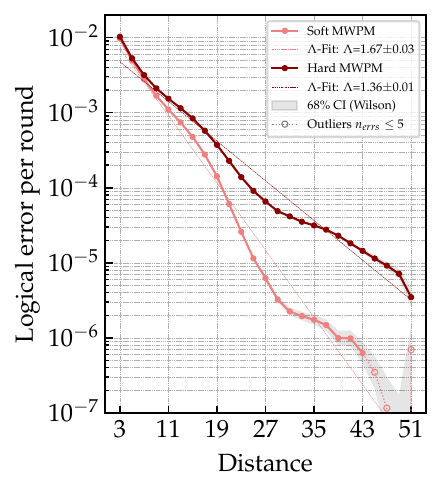}
        \label{fig:RepCodes_soft_X0_lambda}
    }
    \hfill 
    \subfigure[Prepared logical state $\ket{\psi}_L = \ket{-x}_L$.]{
        \includegraphics[width=0.46\columnwidth]{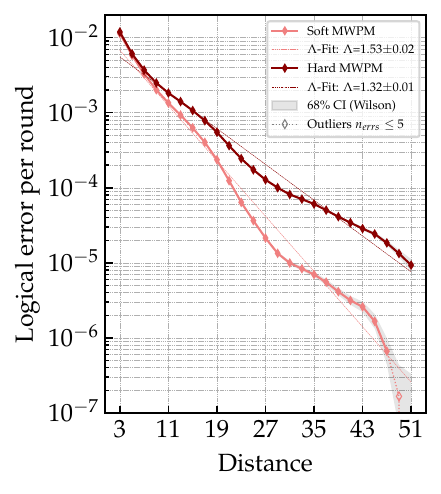}
        \label{fig:RepCodes_soft_X1_lambda}
    }

    \caption{\texttt{IBM Sherbrooke} device data depicting the logical errors per round against code distances for the hard and soft MPWM decoders for $T=50$ rounds of stabilizer measurements. The Lambda factors $\Lambda$ are fitted from the curves' slopes in logarithmic space.}
    \label{fig:HW_eL_combined}
\end{figure}

Fig.~\ref{fig:HW_eL_combined} depicts the logical error per round against code distance for $T=50$ measurement rounds for all four prepared logical states. It also displays the fitted Lambda factor $\Lambda$ for both hard and soft decoding results for every odd distance from $d=3$ to $d=51$, along 68\% confidence intervals using the Wilson score interval method. Similar to the simulation results, points representing fewer than \textit{five} logical error events appear as dashed lines, indicating insufficient data to statistically represent the logical error rate.

The overall trend aligns with our simulations, but several crucial differences emerge. The logical error rates observed are significantly higher than those in the simulations. Comparatively, the observed logical error rates imply that the physical noise is about 1.5 times greater than the calibration data suggests. However, as previously discussed, we attribute this to correlated errors such as crosstalk and leakage. Consequently, drawing a direct connection is challenging since our noise model does not incorporate these correlated errors.

Despite these higher error rates, they maintain a linear relationship with distance on a logarithmic scale, indicative of the expected exponential suppression with increasing code size. Although some curvature was already evident in our simulations, attributed to subsampling, this effect appears to be more pronounced in the hardware data. We suspect that the anisotropy of the hardware noise, which leads to varying error rates across different qubit subsets, amplifies the subsampling curvature effects. This noise variation might also contribute to higher error rates, particularly if specific low-quality qubits disrupt the repetition code chain, drastically reducing code performance as discussed in Appendix \ref{appendix:experimental_setup}. Moreover, the logical error rate curves exhibit some variability depending on the prepared logical state, yet the overall trend remains consistent across all states.

Importantly, the soft decoder consistently outperforms the hard decoder, with this advantage being even more pronounced than in simulations, as will be further addressed in the next section. The soft decoder achieves logical error rates up to 30 times lower than those of the hard decoder, with a magnitude of improvement observed as early as $d=35$ and $d=27$ for prepared states in the $X$ and $Z$ bases, respectively.

This advantage stems from the steeper slopes of the soft decoder compared to the hard decoder, with this difference also being more substantial than in simulations. We present a summary of the Lambda factors in Table \ref{table:lambda_factors_50_rounds}. By taking the ratio of the two lambda factors, we can determine the ratio of the thresholds of the two decoders. This calculation reveals that the threshold of the soft decoder is $24.4\%$ higher than that of the hard decoder when averaged over the four prepared logical states. This enhancement significantly surpasses statistical noise, mirroring the simulations and demonstrating a clear \textit{exponential improvement} of soft decoding for repetition codes implemented on hardware.

\begin{table}
    \centering
    \begin{tabular}{|c|c|c|c|}
        \hline
        \textbf{Logical state} & \textbf{Hard $\mathbf{\Lambda}$-factor} & \textbf{Soft $\mathbf{\Lambda}$-factor}  & \textbf{Increase} \\
        \hline
        $\ket{+x}_L$ & $1.36\pm0.01$ & $1.67\pm0.03$ & $+22.8\%$ \\ \hline
        $\ket{-x}_L$ & $1.32\pm0.01$ & $1.53\pm0.02$ & $+15.9\%$ \\ \hline
        $\ket{+z}_L$ & $1.31\pm0.01$ & $1.63\pm0.02$ & $+24.4\%$ \\ \hline
        $\ket{-z}_L$ & $1.36\pm0.02$ & $1.83\pm0.02$ & $+34.6\%$ \\ \hline
    \end{tabular}
    \caption{Comparison of $\Lambda$-factors for various logical states using hard and soft decoding methods across $T=50$ rounds of stabilizer measurements. The soft decoder shows an average threshold improvement of $24.4\%$. The $\Lambda$-factors are determined from the error rate per round depicted in Fig.~\ref{fig:HW_eL_combined}.}
    \label{table:lambda_factors_50_rounds}
\end{table}

\subsection{Comparison of dynamic and static reweighing}

\begin{figure}
    \centering
    \subfigure[Prepared logical state $\ket{\psi}_L = \ket{+z}_L$.]{
        \includegraphics[width=0.46\columnwidth]{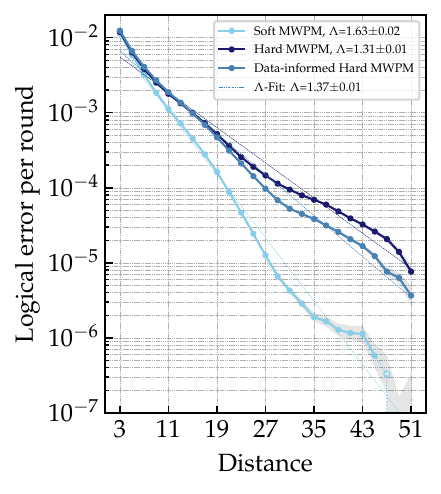}
        \label{fig:MEAN_RepCodes_soft_Z0_lambda}
    }
    \hfill 
    \subfigure[Prepared logical state $\ket{\psi}_L = \ket{-x}_L$.]{
        \includegraphics[width=0.46\columnwidth]{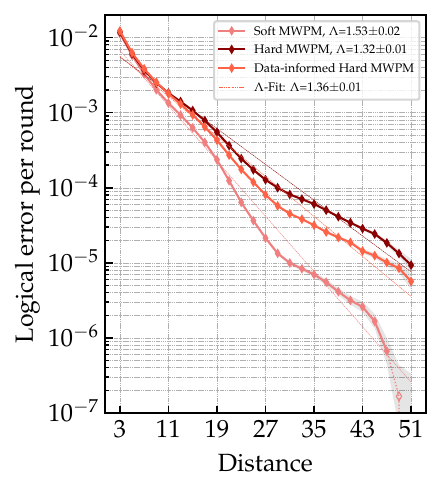}
        \label{fig:MEAN_RepCodes_soft_Z1_lambda}
    }\\
    \caption{Comparison of logical errors per round for data-informed hard decoding, dynamic soft decoding, and calibrated hard decoding, based on experimental data from \texttt{IBM Sherbrooke} for $T=50$ rounds of measurements.}
    \label{fig:MEAN_hardware_e_L_combined}
\end{figure}

Additionally, to verify that the advantages of soft decoding stem from its dynamic reweighting, we decoded the data using the {\it Data-Informed Hard Decoding} method. Similar to Calibrated Hard Decoding, this method employs a non-dynamic approach. Unlike the former, it does not use pre-experiment calibration data for $p_s$; instead, it calculates the mean soft flip probability from all soft outcomes collected throughout the experiment. This approach necessitates statistical data collection before decoding, which leads to complications in a real-time error correction scenario. However, it serves a crucial experimental role by verifying that the value added by dynamically weighting effectively addresses the variance in the ambiguity of measurements on a shot-by-shot basis. Doing so confirms that shot-per-shot adjustments to the weights genuinely enhance the decoding process. Fig.~\ref{fig:MEAN_hardware_e_L_combined} reveals that the data-informed hard decoder outperforms the calibrated hard decoder yet falls short of the soft decoder's effectiveness. This outcome underscores the critical role that dynamic reweighting plays in enhancing decoding with analog information.

\subsection{Decoding with a single byte of soft information}

Soft information decoding offers a promising approach to improve error correction on quantum hardware but requires increased data collection from measurements. This increased data need complicates its practical implementation, particularly in real-time decoding scenarios. To address this issue, we explore the trade-off between the volume of information required from measurements and the resulting performance improvement.

To assess the impact of reducing the precision of the soft flip probabilities on soft decoding, we repeat the simulations from Section \ref{subsection:T50_simulations} with $T=50$ and incrementally increase the accuracy of the soft flip probability from $b=1$ to $b=15$ bits as described in Appendix \ref{appendix:RepCodes_infoPerfo_reduce_information}.

We calculate the ratio of the logical error rate of the decoder with truncated accuracy to that of the decoder with total accuracy. This calculation, represented as ${P_L^b}/{P_L^{64}}$, is plotted against the number of bits $b$ used for the truncation. A ratio of 1 suggests that the decoder with truncated accuracy performs on par with the full-accuracy decoder. This ratio allows us to ``normalize'' the results and, thus, compare multiple distances within the same plot.

\begin{figure}
    \centering
    \includegraphics[width=\columnwidth]{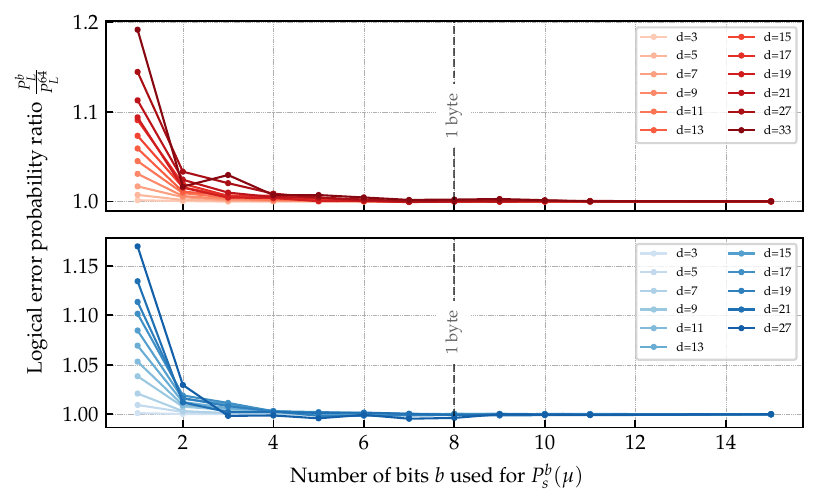}
    \caption{Simulated logical error probability ratio ${P_L^b}/{P_L^{64}}$ for multiple code distances as a function of the number of bits $b$ used to truncate the soft flip probabilities for the states $\ket{+x}$ in red and $\ket{+z}$ in blue. The simulations are performed at twice the physical error rates of \texttt{IBM Sherbrooke}, focusing on code distances $d=33$ for the $X$-basis and $d=27$ for the $Z$-basis to maintain sufficiently high error rates.}
    \label{fig:simul_infoPerfo_T50}
\end{figure}

However, this method requires a substantial number of logical error events to stabilize the logical error rate ratio. Consequently, we use the simulations at twice the device's physical error rates. Additionally, we restrict our analyses to $d=33$ for the $X$-basis and $d=27$ for the $Z$-basis to maintain sufficiently high error rates. We set these limits to align with the benefits observed from soft decoding on hardware, demonstrating improvements exceeding an order of magnitude beyond these thresholds, resulting in highly variable ratios in the device data thereafter. The results of these simulations are presented in Fig.~\ref{fig:simul_infoPerfo_T50}.

Most notably, the data reveals that the ratio ${P_L^b}/{P_L^{64}}$ consistently converges to $1$ across all distances and both bases, resulting in a \textit{convergence point}. Remarkably, this convergence occurs at $b=6$ bits, which indicates that just \textit{one byte} of information per measurement is more than sufficient to match the performance of the full-accuracy decoder.

\begin{figure}
    \centering
    \includegraphics[width=\columnwidth]{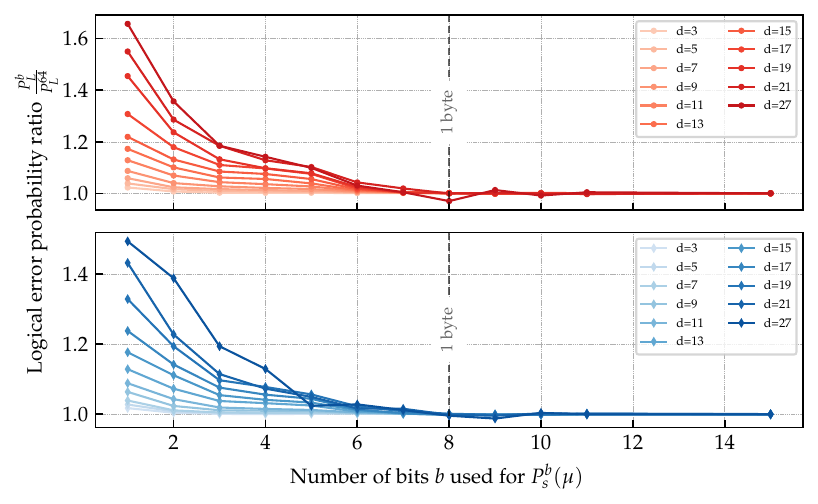}
    \caption{\texttt{IBM Sherbrooke} device data depicting the logical error probability ratio ${P_L^b}/{P_L^{64}}$ for varying code distances as a function of the number of bits $b$ used to truncate the soft flip probabilities. The states $\ket{+x}$ and $\ket{-z}$ are displayed in red and blue, respectively. Distances are limited to $d=33$ for the $X$-basis and $d=27$ for the $Z$-basis to maintain a sufficiently stable probability ratio.}
    \label{fig:HW_infoPerfo_T50}
\end{figure}

To validate the simulation results, we implemented the same methodology on hardware, revisiting experimental data from the previous section with $T=50$ using reduced accuracy soft flip probabilities. The findings, illustrated in Fig.~\ref{fig:HW_infoPerfo_T50}, showcase the logical error rate ratio ${P_L^b}/{P_L^{64}}$ for varying code distances and different bit resolutions $b$. We only show the results for the $|+z\rangle_L$ and the $|-x\rangle_L$ states for compactness.

Additionally, we limited our evaluation to code distances up to $d=27$ for the $X$ and $Z$ basis, maintaining high error rates necessary for stable ratio calculations. We imposed these constraints in line with the observed advantages of soft decoding on hardware in the previous section, where error rates beyond these thresholds were crucially lower, implying significantly variable ratios.

The hardware results mirror the simulation outcomes, with the logical error rate ratio converging to 1 across all distances and states. Again, more considerable distances demonstrate increased sensitivity to the precision of the soft flip probabilities. However, a crucial difference lies in the convergence point, which occurs later at $b=8$ bits.

We hypothesize that the earlier convergence in our simulations results from the diminished role of soft flip probabilities in the overall error budget, given that we simulate at twice the noise rates of the hardware. Exploring the convergence behavior across varying noise rates and devices could be an avenue for future research.

Nevertheless, the crucial behavior---the convergence point across all distances at a specific number of bits---remains consistent between the simulations and the hardware results. This consistency is promising, as it pinpoints a particular number of bits that effectively minimizes the information volume required for soft decoding without compromising performance. In our experiments, this convergence point occurs at one byte, suggesting an \textit{eightfold reduction} in the data volume needed for soft decoding. This reduction underscores the practicality of soft decoding on hardware, including in time-sensitive contexts such as real-time decoding.

More sophisticated truncation strategies could potentially reduce the amount of information even further. However, the presented truncation method is a simple and effective approach that significantly reduces the data footprint while preserving the advantages of soft decoding. Thus, unless specific hardware constraints demand it, implementing more complex strategies for reducing the amount of analog information required for soft decoding may not be necessary.

The consistency in the observed convergence behavior is intriguing and requires further investigation. We suspect a theoretical explanation for this phenomenon, potentially related to the general influence of precision in the weights of decoding graph edges on the performance of MWPM or other graph-based decoders. This topic presents a promising direction for future research.

\subsection{Effect of leakage on the decoding performance}

To demonstrate the effect of leakage in our experiments, we ran repetition codes similar to those presented in Section \ref{subsection:Results_50_rounds} but varied the number of rounds from $T=10$ to $T=100$. Increasing the number of rounds extends the experiment duration, allowing leakage to accumulate. 

\begin{figure}
    \centering
    \subfigure[Soft MWPM decoder.]{
        \includegraphics[width=0.46\columnwidth]{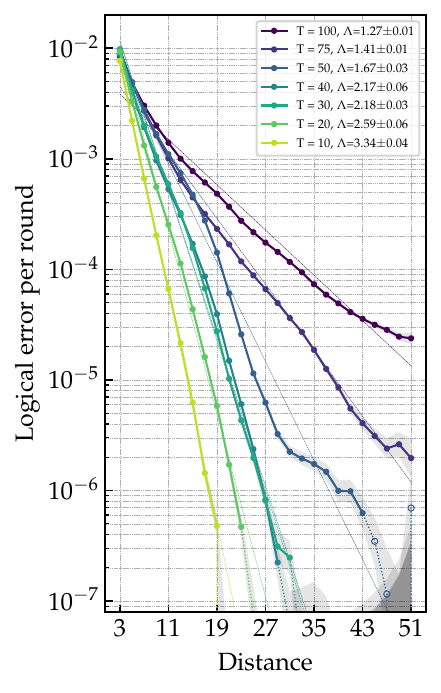}
        \label{fig:e_L_per_rounds_soft}
    }
    \hfill 
    \subfigure[Hard MWPM decoder.]{
        \includegraphics[width=0.46\columnwidth]{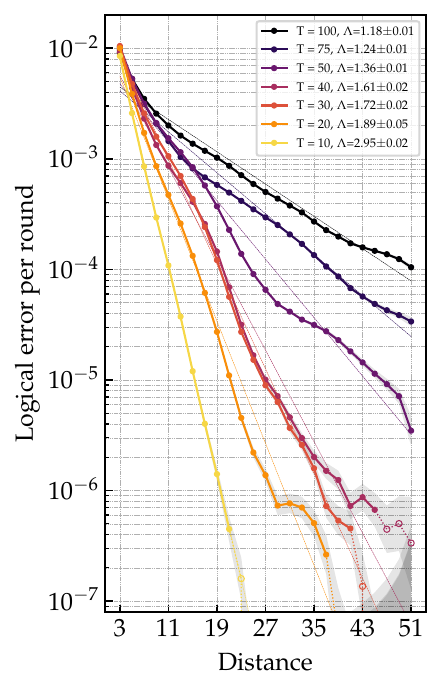}
        \label{fig:e_L_per_rounds_hard}
    }\\
    \caption{\texttt{IBM Sherbrooke} device data depicting the logical error per round plotted against distance using both soft and hard MWPM for varying total round counts $T$. Results are shown for the $\ket{+x}$ state, with consistent behavior across all prepared states.}
    \label{fig:e_L_diff_rounds}
\end{figure}

We plotted the logical error per round $\epsilon_L$ against distance for various total round counts $T$ using both hard and soft decoders in Fig.~\ref{fig:e_L_diff_rounds}. Additionally, we depicted the fitted lambda factors for each round. For compactness, we only show the results for the $\ket{+x}$ state as the behavior is consistent across all prepared states.

The logical error rate per round shows an apparent increase as the number of rounds grows for hard and soft decoders. The growth rate is less pronounced for the soft decoder than for the hard decoder, but it remains considerable. This trend reveals the substantial effect of leakage on both decoding strategies, even when handling leaked measurement outcomes as maximally ambiguous in soft decoding. Moreover, the observed increase in logical error rate is consistent with the findings from Ref.~\cite{kelly_state_2015}, who noted a decrease in the $\Lambda$-factor from round to round when implementing repetition codes on hardware.

\begin{figure}
    \centering
    \includegraphics[width=\columnwidth]{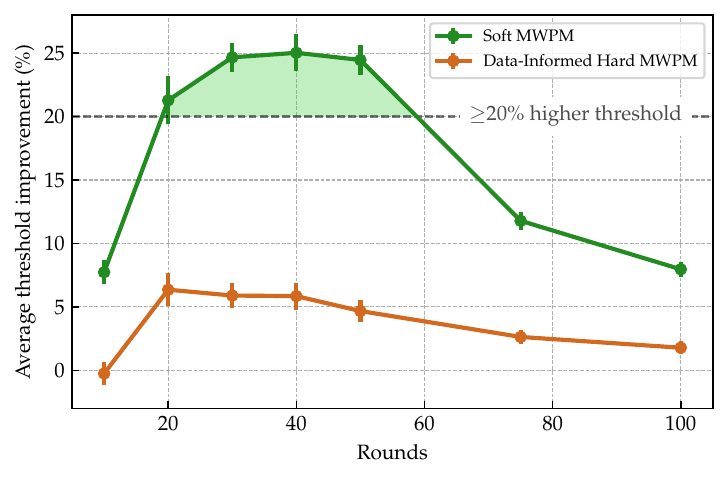}
    \caption{Average threshold improvement of soft and data-informed decoding over hard decoding as a function of the number of rounds. The results are based on experimental data from \texttt{IBM Sherbrooke}.}
    \label{fig:avg_benefit_per_rounds}
\end{figure}

Observing Fig.~\ref{fig:e_L_diff_rounds} and comparing the soft and hard decoders across different numbers of rounds reveals an interesting feature---a variable advantage for soft decoding. To further demonstrate this variable advantage, we calculated the average threshold improvement as a function of rounds in Fig.~\ref{fig:avg_benefit_per_rounds}. Here, we see that the benefit of soft decoding peaks in the mid-range of rounds and diminishes for both lower and higher rounds.

Fig.~\ref{fig:avg_benefit_per_rounds} reveals that the advantage of soft decoding is influenced by the number of rounds and the corresponding amount of leakage. These results indicate that while this leakage management technique does not eliminate the issues associated with leakage, it plays a vital role in the effectiveness of the soft decoding strategy, significantly enhancing its performance. It suggests that our leakage management is, in fact, a primary reason for the substantial improvement of the soft decoder over the hard decoder. This conclusion is supported by the observation that simulations, which lack leakage, show only a 13.6\% improvement in the threshold, whereas hardware results demonstrate a 24.4\% improvement between soft and hard decoding.

We can attribute the exact pattern observed in Fig.~\ref{fig:avg_benefit_per_rounds} to several causes: with fewer rounds, two contributing factors emerge---fewer temporal edges are available for targeted adjustments, and the leakage population is lower, diminishing the soft decoder's advantage in handling leaked states. Conversely, excessive rounds increase leakage, resulting in significant errors affecting the qubits and gates that interact with leaked qubits. This increase in errors dilutes the benefits of managing measurement ambiguities, as they become overshadowed by more prevalent errors elsewhere in the system. Furthermore, given the suboptimal nature of our leakage detection process, the accuracy of the soft decoder diminishes as leakage intensifies, hence further contributing to the diminishing benefits of soft decoding at higher rounds.

Similarly, the data-informed hard decoder shows near-zero benefits for both high and low rounds but maximizes in mid-rounds. In low rounds, leakage is minimal, so the mean ambiguity of the IQ points remains close to that obtained from calibration data. In mid-rounds, more leakage occurs, and informing the decoder about the generally higher uncertainty of our measurement points aids in handling this leakage---though not as effectively as individually indicating the ambiguity for each measurement outcome. However, for many rounds, extensive leakage affects surrounding qubits and gates, thus diminishing the benefits of informing the decoder about the mean ambiguity of measurement edges compared to mid-rounds.

\section{Discussions}

Our results show that soft decoding offers a clear advantage over hard decoding. However, a significant portion of the observed advantage on hardware arises from our approach to managing leakage as maximally ambiguous rather than handling the ambiguity of $\ket{0}$-$\ket{1}$ state discrimination.

Our simulations of the repetition code experiments emphasize the role of leakage in the benefit of soft decoding. In retrospect, the simulations can be seen as a tool to investigate the theoretical leakage-free behavior of soft decoding, as we did not include leakage in our noise model. They suggest that the advantage of using soft information should be at an average threshold increase of about 13\%. However, our hardware experiments show that the advantage of soft decoding is significantly greater with a 24.4\% higher threshold. This discrepancy underlines the role of treating the leaked states as maximally ambiguous.

The impact of leakage on the soft decoding's advantage is further underlined by the threshold improvements when running experiments with varying numbers of rounds and, thus, leakage populations. We observe that the benefit of soft decoding peaks in the mid-range of rounds with more than a 20\% higher threshold and decreases for both lower and higher rounds to around a 7\% improvement. This pattern results from the reduced leakage with a lower number of rounds, diminishing the soft decoder's advantage in handling leaked states. Conversely, with a higher number of rounds, increased leakage leads to widespread errors throughout the system, undermining the advantage of managing measurement ambiguities and thus lessening the soft decoder's benefit.

The importance of our proposed leakage treatment for the benefit of soft decoding raises the question of whether a more refined approach to leakage handling could yield even more significant benefits. 

Nevertheless, we believe that our technique is a straightforward and efficient way to improve the performance of soft decoding in scenarios involving leakage. However, it is not a universal remedy, particularly in cases of severe leakage. Therefore, addressing significant leakage through additional mitigation strategies is advisable. Once large-scale leakage is controlled, soft decoding can be effectively applied to manage any residual leakage. This strategy ensures that our adapted soft decoding is employed where it is most advantageous, enhancing its benefits compared to hard decoding.

\section{Conclusions}

Our hardware experiments demonstrate a substantial benefit of soft decoding, resulting in up to 30 times lower logical error rates due to a 24.4\% higher threshold of the soft decoder compared to the hard decoder. This advantage primarily stems from dynamically reweighting the decoding graph based on the analog measurement outcomes since static decoding with data-informed weights does not provide comparable benefits.

While soft decoding requires handling more data than hard decoding, our findings suggest that we can reduce the precision of the soft flip probabilities to just one byte without compromising the decoder's performance. This data-efficient decoder facilitates the application of soft-information decoding in real-time decoding scenarios. If needed by hardware requirements, we can further reduce the precision with tailored reduction schemes.

Finally, we observe that the benefit of soft over hard decoding in our device data is largely due to the presence of leakage and our method of treating the detected leakage as maximally ambiguous measurement outcomes. This method is particularly effective in scenarios with moderate leakage, specifically for up to approximately $T\approx50$ rounds of stabilizer measurements. In this regime, our leakage handling technique could complement hardware-based leakage reduction protocols, reducing the frequency with which leakage reduction is applied. Such a combined approach would decrease the error budget stemming from leakage reduction while maintaining appropriate error suppression.

\section{Acknowledgements}


The authors thank I. Hesner, M. Beverland, E. Chen, A. Cross, E. Pritchett, M. Carroll, and M. Li for their valuable discussions and insights.

This research was sponsored by the Army Research Office and was accomplished under Grant Number W911NF-21-1-0002. The views and conclusions contained in this document are those of the authors and should not be interpreted as representing the official policies, either expressed or implied, of the Army Research Office or the U.S. Government. The U.S. Government is authorized to reproduce and distribute reprints for Government purposes notwithstanding any copyright notation herein.

BH acknowledges support from the NCCR SPIN, a National Centre of Competence in Research, funded by the Swiss National Science Foundation (grant number 51NF40-180604).

\section{Data and Code Availability}

The code used for the experiments and analysis in this paper is available at \href{https://github.com/quantumjim/soft_information_decoding}{github.com/quantumjim/soft-information-decoding}. The data supporting the findings of this study will be made available upon reasonable request.

\clearpage

\appendix

\section{Soft information in other quantum computing platforms}

Other quantum computing technologies face similar challenges in achieving accurate qubit state discrimination. For instance, the state measurement in trapped ions involves detecting fluorescence from ions when illuminated with a laser. Ideally, ions in the excited state fluoresce while those in the ground state do not. However, due to background photon count, laser intensity fluctuations, and photodetector efficiency, the resulting signal also forms a distribution of counts. The overlap between distributions for \textit{dark state} (ground state) and \textit{bright state} (excited state) similarly leads to a finite misassignment probability~\cite{bruzewicz_trapped-ion_2019}. 

In semiconductor quantum dot systems, qubit state readout is often performed through charge sensing, where the tunneling of electrons to and from the quantum dot affects the conductance of a nearby sensor. The conductance signal, influenced by the qubit's charge state, generates distributions that must be distinguished to infer the qubit state accurately. As with superconducting and ion trap platforms, variations in tunnel rates, sensor sensitivity, and environmental noise contribute to the overlap between these distributions, posing a challenge for reliable qubit state discrimination~\cite{burkard_semiconductor_2023}.

These examples underscore a common theme across quantum computing platforms: the necessity of managing and interpreting probabilistic signals derived from physical phenomena unique to each technology. The goal remains to optimize signal-to-noise ratios and separation between state-indicative distributions, minimizing the likelihood of misassignment errors.

\section{Dynamic weighting algorithm}

The following algorithm dynamically weights the decoding graph based on the soft outcomes of stabilizer measurements. It calculates the final error probability of code qubits with
\begin{align}
    \label{eq:total_error_probability_code_channels}
    \mathbb{P}_E[\mathtt{E_c^{tot}}] = & \ \  p_e(1-p_s)(1-p_h) \nonumber \\ \nonumber&+(1-p_e)p_s(1-p_h) \\\nonumber&+(1-p_e)(1-p_s)p_h \\&+  p_e p_s p_h.
\end{align}
where $p_s$ ($p_h$) are soft (hard) flip probabilities and $p_e = \mathbb{P}_E[\mathtt{E_c}]$ is the probability of an error $\mathtt{E}_c$ happening on a code qubit $c$ due to gate, idling, or propagation errors. 

\begin{algorithm}[H] 
    \caption{Dynamic Reweighting of Soft Edges in the Decoding Graph}
    \label{algorithm:soft_decoding}
    \begin{algorithmic}[1]
        \Statex \textbf{Input:}
        \State Weighted decoding graph $G_T$ for $T$ rounds of measurements, with stabilizer nodes $\{n_{a, t} \mid a\in\{1,\ldots,m\}, t\in\{1,\ldots,T\}\}$.
        \State Soft stabilizer measurement outcomes $\{\boldsymbol{\mu}_t\}_{t=1,\ldots,T}$.
        \State Soft code measurement outcomes $\mu^{c_1}, \ldots, \mu^{c_n}$.
        \State Probability distributions $f^{(0)}(\mu)$ and $f^{(1)}(\mu)$.
        \Statex \textbf{Output:}
        \State A reweighted decoding graph $\tilde{G}_T$.
        
        \For{\textbf{each} $t = 1$ \textbf{to} $T$}
            \For{\textbf{each} stabilizer $a = 1$ \textbf{to} $m$}
                \State Retrieve the soft outcome $\ensuremath{\mu_{a,t}}$ from $\ensuremath{\boldsymbol{\mu}_t}$.
                \State Compute the soft flip probability \ensuremath{p_s(\mu_{a,t})} using Equation \ref{eq:soft_flip_probability}.
                \If{$t \leq T-1$}
                    \State Calculate the edge weight \ensuremath{w = -\log(p_s / (1-p_s))}.
                    \State Reweight the distance-two edge \ensuremath{(n_{a,t}, n_{a,t+2})}.
                \EndIf
                \If{$t = T$}
                    \State Compute the total measurement error probability using Equation \ref{eq:soft_and_hard_probability_last_round}.
                    \State Calculate the edge weight \ensuremath{w = -\log(p_{\text{tot}} / (1-p_{\text{tot}}))}.
                    \State Reweight the distance-one edge \ensuremath{(n_{a,t}, n_{a,t+1})}.
                \EndIf
            \EndFor
        \EndFor
        
        \For{\textbf{each} code outcome $\mu^{c_i}$}
            \State Identify the error channel involving code qubit \ensuremath{c_i}.
            \State Compute the soft flip probability \ensuremath{p_s(\mu^{c_i})} using Equation \ref{eq:soft_flip_probability}.
            \State Compute the total error probability of that channel using Equation \ref{eq:total_error_probability_code_channels}.
            \State Calculate the edge weight \ensuremath{w = -\log(p_{\text{tot}} / (1-p_{\text{tot}}))}.
            \State Reweight the edge corresponding to the identified error channel.
        \EndFor
    \end{algorithmic}
\end{algorithm}

\section{\label{appendix:noise_model}Noise model estimation}

We estimate our noise model using per-qubit calibration data from IBM Quantum devices, which include single and two-qubit gate fidelities and the $T_1$ and $T_2$ decoherence times. These times are used in 
\begin{equation}
        \label{eq:idling_error_probabilities_p_X_P_Y}
        p_X = p_Y = \frac{1-e^{-\tau / T_1}}{4},
\end{equation}
\begin{equation}
        \label{eq:idling_error_probabilities_p_Z}
        p_Z=\frac{1-e^{-\tau / T_2}}{2}-\frac{1-e^{-\tau / T_1}}{4}.  
\end{equation}
to estimate the idling error probabilities. Specifically, we set the idling time $\tau$ in each round to correspond to the average measurement time of the ancilla qubits, as the code qubits are only idle while the ancilla qubits get read out. These calibrations occur roughly once per day. 

\begin{figure}
    \centering
    \includegraphics[width=0.7\columnwidth]{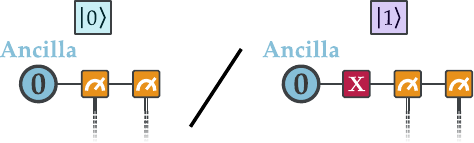}
    \caption{Double measurement calibration circuit for estimating the probabilities of soft and hard flips in the first measurement.}
    \label{fig:calibration_circuit}
\end{figure}

We run a set of calibration circuits before each experiment to estimate the probabilities for the measurement-related processes. These circuits consist of preparing all the device qubits in the $|0\rangle$ or $|1\rangle$ state and measuring them twice in the $Z$ basis, as depicted in Fig.~\ref{fig:calibration_circuit}. The second measurement allows us to differentiate between soft and hard flips by comparing the two measurement outcomes. For example, preparing the $|1\rangle$ state, we summarize the possible estimated outcomes $\hat{z}$ and contributing error channels in Table \ref{table:calibration_circuits}. We assume that the reset error rate and the single $X$ gate error rate are negligible compared to the measurement error rates.

\renewcommand{\arraystretch}{2} 
\begin{table}[!ht]
    \centering
    \begin{tabular}{|c|c|c|c|}
        \hline
        \textbf{Outcomes} & \textbf{Smallest order} & \textbf{Second smallest order}  \\
        \hline
        $\mathtt{11}$ & $\emptyset^\mathtt{1}\emptyset^\mathtt{2}=\mathcal{O}(p^0)$\ & $\mathtt{E_s^1E_s^2}=\mathcal{O}(p^2)$ \\ \hline
        $\mathtt{00}$ & $\mathtt{E_h^1\emptyset^2}=\mathcal{O}(p^1)$ & $\mathtt{E_s^1E_h^2} + \mathtt{E_s^1E_s^2}=\mathcal{O}(p^2)$  \\ \hline
        $\mathtt{10}$ & $\mathtt{\emptyset^1E_h^2} + \mathtt{\emptyset^1E_s^2} = \mathcal{O}(p^1)$ & $\mathtt{E_h^1E_s^1} + \mathtt{E_h^1E_s^2} = \mathcal{O}(p^2)$ \\ \hline
        $\mathtt{01}$ & $\mathtt{E_s^1\emptyset^2} = \mathcal{O}(p^1)$ & $\mathtt{E_h^1E_h^2} + \mathtt{E_h^1E_s^2} = \mathcal{O}(p^2)$ \\
        \hline
    \end{tabular}
    \caption{Summary of the possible outcomes and contributing error channels when preparing a $|1\rangle$ state and measuring it twice. $\emptyset^\mathtt{m}$, $\mathtt{E_h^m}$, $\mathtt{E_s^m}$ denotes the absence of an error, a hard flip, and a soft flip on measurement $m$, respectively.}
    \label{table:calibration_circuits}
\end{table}

Ignoring higher order errors $\mathcal{O}(p^2)$, a \textit{soft flip} occurs on the first measurement when the outcome is $\mathtt{0}$ and the second outcome is $\mathtt{1}$ again. We can interpret these outcomes as having had a misalignment on the first measurement, but the state was unchanged, leading to a correct outcome on the second measurement. On the other hand, a \textit{hard flip} occurs when the first outcome is $\mathtt{0}$ and the second outcome is $\mathtt{0}$ again. This outcome means the qubit decayed to the $|0\rangle$ state during the first measurement and was then measured correctly. 

Using the calibration data, we can estimate the probabilities of soft and hard flips in the first measurement for the $|1\rangle$ state for each qubit as
\begin{equation}
    p_s = \frac{N_\mathtt{01}}{N_{\text{tot}}}, \quad p_h = \frac{N_\mathtt{00}}{N_{\text{tot}}},
\end{equation}
with $N_{\mathtt{ij}}$ the number of times the outcomes $\mathtt{ij}$ were observed and $N_{\text{tot}} = N_\mathtt{01}+N_\mathtt{00}+N_\mathtt{10}+N_\mathtt{11}$ the total number of measurements. Accordingly, the probabilities for the $|0\rangle$ state are
\begin{equation}
    p_s = \frac{N_\mathtt{10}}{N_{\text{tot}}}, \quad p_h = \frac{N_\mathtt{11}}{N_{\text{tot}}}.
\end{equation}
These probabilities represent the \textit{mean probabilities} of soft and hard flips and are used for the \textit{Calibrated Hard Decoding} discussed in Section \ref{subsection:Results_50_rounds}.

Our experiments will average the probabilities $p_s$ and $p_h$ and the gate and idling errors over all the qubits involved. By averaging, we can mitigate inaccuracies due to potential noise drift.

\section{\label{subsection:kernel_density_estimation}\texorpdfstring{Kernel Density Estimation of $f^{(\bar{z})}(\mu)$}{Kernel Density Estimation of f(z)(mu)}}

\begin{figure}
    \centering
    \includegraphics[width=0.7\columnwidth]{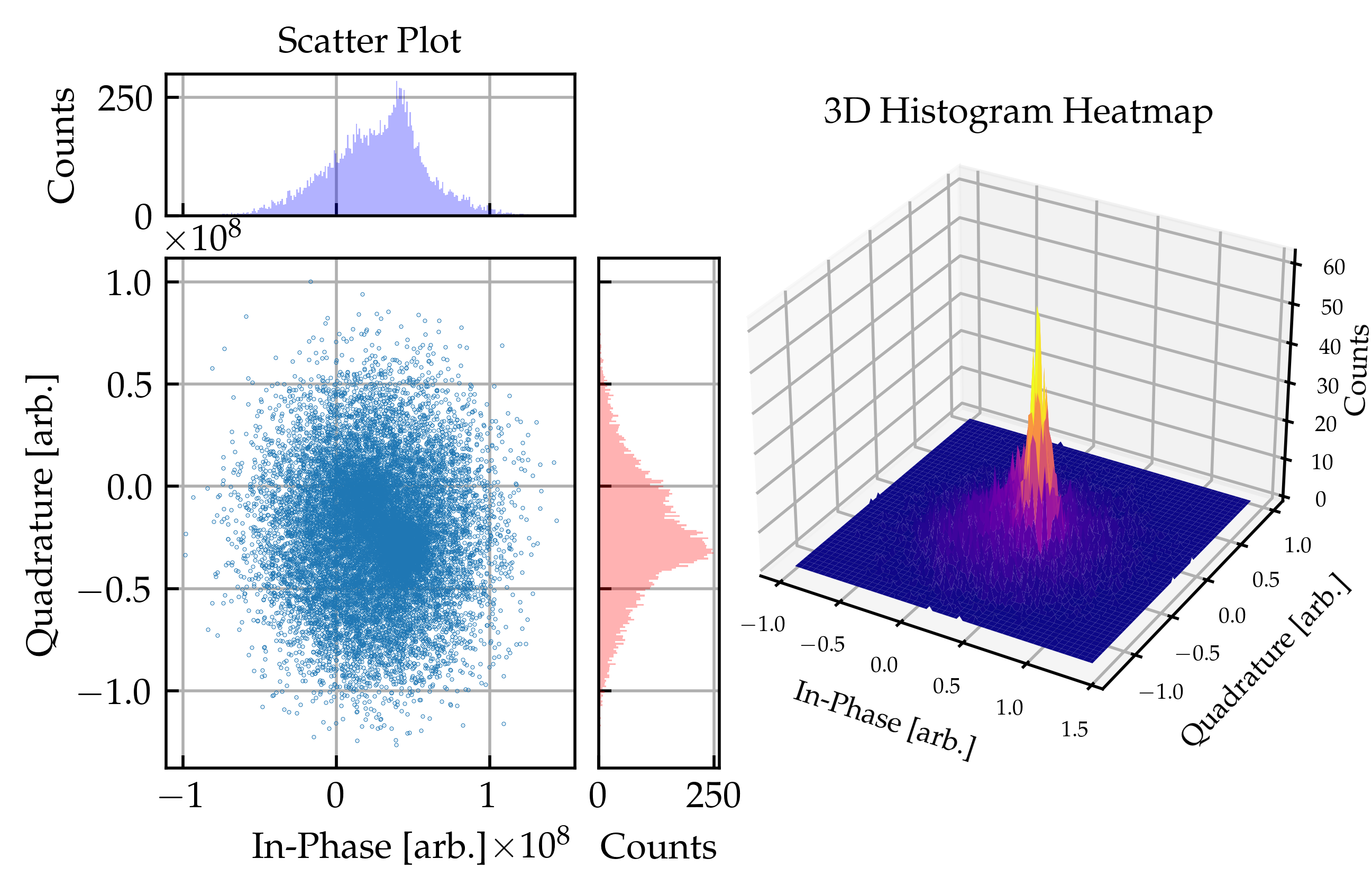}
    \caption{IQ data for a particularly bad-mannered qubit (106) of the \texttt{IBM Sherbrooke} device prepared in the $\ket{1}$ state. The data shows a complex behavior that is not easily modeled by a simple Gaussian distribution.}
    \label{fig:iq_data_bad_mannered}
\end{figure}

To estimate the distributions $f^{(0)}(\mu)$ and $f^{(1)}(\mu)$ we run the calibration circuits, retrieve the soft outcomes $\mu$, and then fit the distributions of these points using \textit{Kernel Density Estimation}. Kernel density estimation is a non-parametric method for estimating the probability density function of a random variable. We prefer this method over parametric methods like a \textit{Gaussian Mixture Model} as it does not require assumptions about the underlying distribution. Moreover, as seen in Fig.~\ref{fig:iq_data_bad_mannered}, the distributions of the soft outcomes comprise complex shapes that could lead to inaccuracies when using Gaussian mixture models. 


KDE is based on the convolution of a kernel function with the data points. We use the \textit{Epanechnikov kernel} function as it has a few advantages. Statistically, it is the optimal kernel because it provides the best trade-off between bias and variance. Moreover, it has a compact support, meaning that its value is zero beyond a specific range (\(|u| > 1\)). This property reduces the computational complexity compared to other kernels like the Gaussian kernel since it only accounts for points within its support~\cite{epanechnikov_non-parametric_1969}. The only free parameter is the kernel bandwidth, which determines the smoothness of the estimated distribution, with a smaller bandwidth leading to a more jagged distribution and a larger bandwidth leading to a smoother distribution. We set the bandwidth for each qubit using \textit{cross-validation}, splitting the data into a training and a validation set and choosing the bandwidth that minimizes the validation error.

To estimate the distributions $f^{(0)}(\mu)$ and $f^{(1)}(\mu)$, we run calibration circuits as shown in Fig.~\ref{fig:calibration_circuit}. We use the second measurement outcome to filter out data points where a hard flip occurred, as described in Table \ref{table:calibration_circuits}. This step ensures that we fit the distributions of the qubit's ``true'' initial state rather than its state after the measurement.

However, a complication arises due to the state changing during the readout circuit, meaning the outcome $\mu$ would not correspond to the expected prepared state. To circumvent this issue, we use the second measurement of our calibration circuits depicted in Fig.~\ref{fig:calibration_circuit} to \textit{filter out} these hard flips. Consequently, we run calibrations for the $\bar{z}=|0\rangle$ and $\bar{z}=|1\rangle$ states and remove every shot where the second measurement outcome disagrees with the prepared state. We then retrieve all soft outcomes from the corresponding state $\boldsymbol{\mu}^{\bar{z}}$ and fit the distribution $f^{(\bar{z})}(\mu)$ to these points using kernel density estimation. 

\section{\label{appendix:experimental_setup}Experimental setup}

We implement various configurations of the repetition code, encoding both in the $X$ and $Z$ bases and varying the prepared logical state between $\ket{+z}_L$ and $\ket{-z}_L$. We encode the logical state, measure $T$ rounds of stabilizer measurements, and conclude with a final readout of all code qubits in the same basis as initially encoded.  We inform our noise model with calibration data, and we utilize Pauli tracing via the Stim software to weigh the edges of the decoding graph~\cite{gidney_stim_2021}.

We decode the repetition codes using the MWPM algorithm via the PyMatching software~\cite{higgott_pymatching_2021}. This algorithm determines the most likely error (MLE) by identifying the minimum-weight perfect matching within the decoding graph. To evaluate the logical error rate, we select the last qubit in the chain---though this selection is arbitrary---to carry the logical information. If the MLE affects this logical qubit, we classify it as a logical error. The results from this process allow us to compare the predicted logical error against the actual logical error observed during the final measurement of the code qubit.

\begin{figure}
    \centering
    \includegraphics[width=\columnwidth]{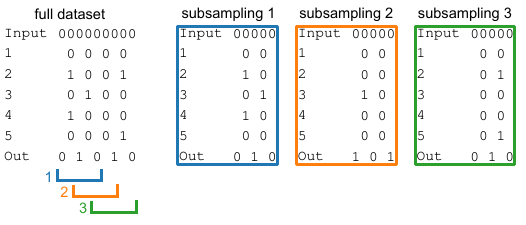}
    \caption{Example of subsampling for a repetition code of distance $d_s=3$ from a $d=5$ code. The subsampling method allows us to evaluate the performance of repetition codes across multiple distances without the need to individually execute each configuration on the hardware. Figure taken from~\cite{google_quantum_ai_exponential_2021}.}
    \label{fig:subsampling}
\end{figure}

We use a technique known as \textit{subsampling}~\cite{kelly_state_2015, naveh_theoretical_2018, google_quantum_ai_exponential_2021} to run repetition codes sample-efficiently on hardware. This approach involves deploying the largest possible repetition code that the hardware can accommodate and decoding various subsets to gather data of smaller distances. This strategy generates $n=d-d_s + 1$ distinct datasets, each corresponding to a repetition code of distance $d_s$. An example of subsampling for a repetition code of distance $d_s=3$ from a $d=5$ is shown in Fig.~\ref{fig:subsampling}. By subsampling, we can analyze the performance of repetition codes across multiple distances without the need to individually execute each configuration on the hardware. This method also ensures consistent noise conditions across different experiments, eliminating noise drifts.

Since repetition codes are linear, they are inherently more susceptible to the impact of individual defective qubits than two-dimensional codes. A malfunctioning ancilla qubit, for instance, could disrupt the integrity of the code, effectively breaking the chain and potentially halving the code's distance. Subsampling smaller distances and averaging over them allows us to systematically evaluate every part of the qubit chain, ensuring that even smaller distances encounter any problematic qubits. However, reducing the distance has an exponential impact on the logical error rate; thus, averaging over subsets does not fully counteract the detrimental effects of bad qubits. Furthermore, as we conclude each experiment with a final code readout, we cannot subsample the number of rounds as simply as we can for the distance. Consequently, we will run different numbers of rounds and use these variations to estimate the logical error rate per round individually.

\section{Simulation setup}

To simulate soft decoding of repetition codes as if we run it on the hardware, we first generate outcomes for stabilizer and final code measurements using the circuit-level Pauli noise model. We configure the circuit level noise model with the calibration data obtained from the hardware for two-qubit gates, single-qubit gates, and decoherence times. We use these values alongside the explicit code circuit to estimate the noise model. The Stim software~\cite{gidney_stim_2021} facilitates the outcome sampling process by providing tools for Pauli tracing, which we also use to weight the graph.

To then simulate the measurement process and get the soft outcomes $\mu$, we use measurement calibrations run on the hardware to estimate and model the density distributions $f^{(0)}(\mu)$ and $f^{(1)}(\mu)$ using kernel density estimation. For each outcome $z$, we sample an IQ point according to the fitted distribution $f^{(z)}(\mu)$ and assign this IQ point to the measurement outcome. We use calibration data from the $\texttt{IBM Sherbrooke}$ device for both the noise model and the IQ distributions. We replicate this procedure for all measurement outcomes to assemble the IQ points for the stabilizer and final code measurements. To ensure that the characteristics of the IQ distributions are the only source of soft flips, we set the probability of a soft flip in our noise model used for the generation of hard outcomes to $p_s = 0$. 

With these IQ points $\mu$, we re-estimate the measurement outcomes $\hat{z}(\mu)$ based on Equation \ref{eq:maximum_likelihood_assignment} and calculate the corresponding soft flip probabilities $p_s(\mu)$ according to Equation \ref{eq:soft_flip_probability}. This re-estimation incorporates soft flips as they would occur on the device. We use the soft flip probabilities to adjust the weights of the decoding graph according to Algorithm \ref{algorithm:soft_decoding}. Subsequently, we decode the outcomes and the resulting syndromes using PyMatching.

The software that implements the estimation of outcomes and soft flip probabilities, as well as the software for reweighting the decoding graph, will be made available soon. To ensure rapid execution times, we wrote the performance-critical components in C++. We achieve the kernel density estimation estimation using the MLPack library~\cite{curtin_mlpack_2023}. 

Importantly, as discussed in Section \ref{subsection:leakage}, the calibration data does not reflect leakage. Consequently, the simulation result fails to capture the full extent of the measurement process that would occur for a more significant number of rounds. However, the simulation results provide valuable insight into the potential benefits of soft decoding repetition codes on hardware.

\section{\label{appendix:RepCodes_infoPerfo_reduce_information}How to reduce the amount of measurement information}

Several possibilities exist to reduce the data footprint in the soft decoding process. However, we expect that reducing the amount of analog information leads to similar performance reductions across different methods. Consequently, we focus on a simple, hardware-inspired approach to minimize data usage. This approach involves estimating the outcome $\hat{z}(\mu)$ and the probability of a misassignment $p_s(\mu)$ from the IQ point $\mu$ directly on the control card. After estimations, the card truncates $p_s(\mu)$ to a specified precision, reducing the number of bits required for its representation. Following this, the control card transmits both the outcome and the truncated probability of a soft flip to the RAM for subsequent decoding. This method effectively reduces the data retrieved from the measurements by processing it at the control card level before only passing the outcome and the truncated probability of a soft flip to the RAM. 

To facilitate fast probability estimation, $f^{(0)}(\cdot)$ and $f^{(1)}(\cdot)$ are pre-estimated during device calibration and stored in a lookup grid, allowing for the computation of the needed $f^{(0)}(\mu)$ and $f^{(1)}(\mu)$ for each IQ point $\mu$ in $\mathcal{O}(1)$ time. Additionally, with multiple control cards, each managing a set of qubits, this estimation process can be parallelized, significantly reducing the computational load on the decoder unit by distributing the workload across multiple control cards.

\begin{figure}
    \centering
    \includegraphics[width=\columnwidth]{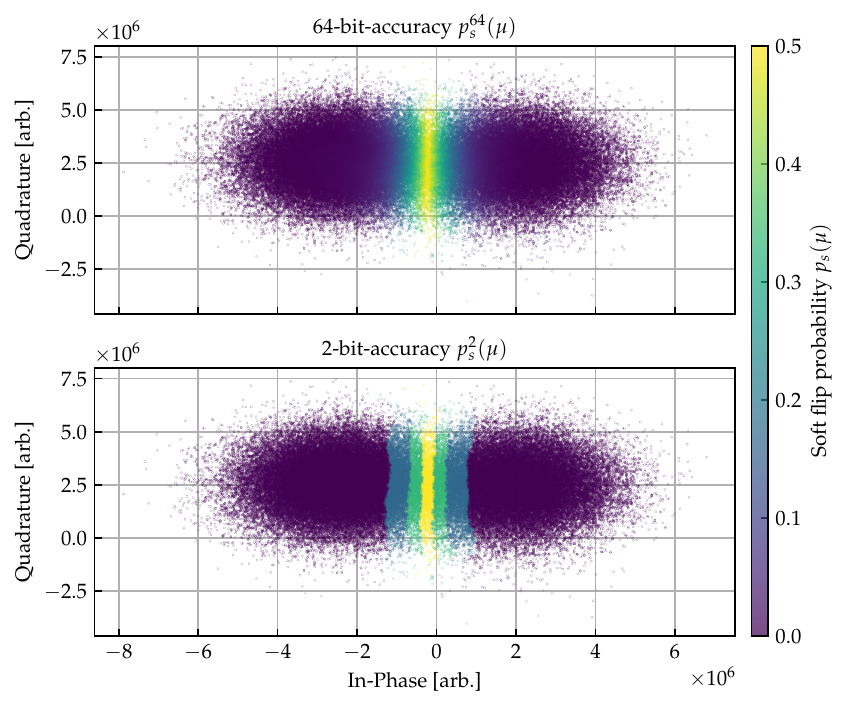}
    \caption{Colormap representation of measurement outcomes $\mu$ for qubit 72 of \texttt{IBM Shebrooke}. The top row shows the full accuracy soft flip probability $p_s^{64}$, while the bottom row illustrates the effects of truncating the soft flip probability to 2 bits.}
    \label{fig:cmap_pSoft_truncation}
\end{figure}

\begin{figure}
    \centering
    \vspace{1em}
    \includegraphics[width=\columnwidth]{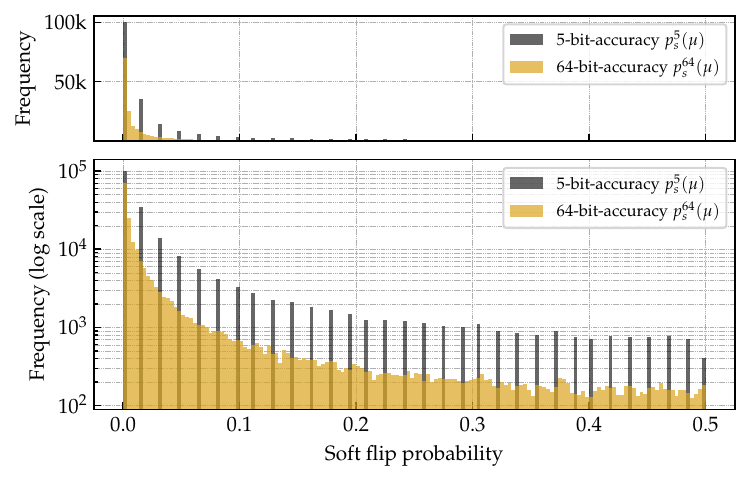}
    \caption{Histogram of soft flip probabilities for qubit 72 in \texttt{IBM Shebrooke}. The plot illustrates the distribution of soft flip probabilities and the effects of truncating the probabilities to 5 bits. Plotting the frequency without a logarithmic scale reveals that a significant portion of points cluster within 0\% and 5\%.}
    \label{fig:histogram_pSoft}
\end{figure}

To assess the impact of accuracy loss due to the truncation of $p_s$, we analyze the decoder's performance using the full accuracy soft flip probability, represented as a 64-bit double ($p_s^{64}$) and compare it to the performance using the truncated soft flip probability ($p_s^{\text{trunc}}$). We achieve this truncation by rounding the double to $b$ bits. An illustration of the effects of truncating $p_s(\mu)$ is presented in Fig.~\ref{fig:cmap_pSoft_truncation} and Fig.~\ref{fig:histogram_pSoft}.

Truncating the soft flip probability $p_s$ uniformly across its entire range is straightforward but may not be the most efficient. It is possible to consider adaptive methods that provide higher precision for ranges where soft flip probabilities are more commonly observed and lower precision elsewhere. As illustrated in Fig.~\ref{fig:histogram_pSoft}, a significant number of points cluster within a specific interval, suggesting that increased accuracy in these areas could enhance efficiency. Additionally, identifying where the decoder is most sensitive to the precision of soft flip probabilities and prioritizing these ranges could further optimize performance. However, as discussed in the subsequent section, this primary truncation method significantly reduces data needs while preserving the advantages of soft decoding. Therefore, unless specific hardware constraints demand it, implementing more complex strategies for reducing the amount of analog information required for soft decoding may not be necessary.

\bibliography{references}

\end{document}